\documentstyle[12pt,epsfig]{article}
\def\IJMP #1 #2 #3 {{\it Int.\ J.\ Mod.\ Phys.}\ {\bf #1}\ (#2) #3}
\def\MPL #1 #2 #3 {{\it Mod.\ Phys.\ Lett.}\ {\bf #1}\ (#2) #3}
\def\NC #1 #2 #3 {{\it Nuovo Cim.}\ {\bf #1} (#2) #3}
\def\NP #1 #2 #3 {{\it Nucl.\ Phys.}\ {\bf #1}\ (#2) #3}
\def\PL #1 #2 #3 {{\it Phys.\ Lett.}\ {\bf #1}\ (#2) #3}
\def\PR #1 #2 #3 {{\it Phys.\ Rev.}\ {\bf #1}\ (#2) #3}
\def\PP #1 #2 #3 {{\it Phys.\ Rep.}\ {\bf #1}\ (#2) #3}
\def\PRL #1 #2 #3 {{\it Phys.\ Rev.\ Lett.}\ {\bf #1}\ (#2) #3}
\def\RMP #1 #2 #3 {{\it Rev.\ Mod.\ Phys.}\ {\bf #1}\ (#2) #3}
\def\ZP #1 #2 #3 {{\it Z.\ Phys.}\ {\bf #1}\ (#2) #3}
\def\g{\gamma}
\begin{document}
\begin{flushright}
JINR E2-96-31\\
hep-ph/9601384
\end{flushright}
\vspace*{1cm}

\begin{center}
{\bf QCD CORRECTIONS TO HEAVY QUARK PAIR PRODUCTION IN POLARIZED
$\g\g$ COLLISIONS AND THE INTERMEDIATE MASS HIGGS SIGNAL}\\

\vspace*{5mm}
{G.~Jikia}\\
{\it Institute for High Energy Physics}\\
{\it Protvino, Moscow region, 142284, Russia}\\
\vspace*{5mm}
{A.~Tkabladze}\\
{\it Bogoliubov Laboratory of Theoretical Physics,}\\
{\it JINR, Dubna, Moscow Region, 141980, Russia}\\
\end{center}

\begin{abstract}
Perturbative QCD one-loop corrections to the cross sections of the $b\bar b
(c\bar c)$ quark pair production in polarized photon-photon collisions, as well
as cross sections of the radiative processes $\g\g\to b\bar b g,\ c\bar c g$
leading to two- and three-jet final states are calculated. It is shown that the
signal from the intermediate mass Higgs boson is observable and precise
measurements of the Higgs boson two-photon width are possible at a
photon-photon collider, although the statisti\-cal significance 
is substantially
reduced with respect to the tree level calculations. We demonstrate that
virtual corrections are of the same order or larger than the Born contribution
in the $J_z=0$ channel at high energy for small values of the cutoff $y_{cut}$,
separating two-jet from three-jet topologies. The nature of these large
corrections is elucidated. For $b\bar b$ pair production at
$\sqrt{s_{\g\g}}\sim 100$~GeV and for small values of $y_{cut}\leq 0.04$ the
higher order resummation of double logarithmic terms ${\cal O}\left(\alpha_s
m_b^2/s\ln^2(s/m_b^2)\right)$ is needed.
\end{abstract}

\section{Introduction}
\setcounter{equation}{0}

Higgs boson search and study of the electroweak symmetry breaking mechanism
will be major goals for the next generation of supercolliders\cite{j1,j2}.
Despite  excellent successes of the standard model (SM) in  describing of
experimental data of electroweak interactions of gauge bosons and fermions, no
evidence, even indirect, for the Higgs boson sector of the theory has been
given yet. Much has been written on the search for the SM Higgs boson at Hadron
Supercolliders (see e.g. ref \cite{j2} for recent reviews and further
references).

A Higgs particle with mass above  $2 m_W$ ($2 m_Z$) can be produced and
detected in its $W^+W^-$ or $ZZ$ decay mode at high-energy hadron colliders
because the continuum of massive vector boson pair production is significantly
small. It is generally believed that a  Higgs boson of mass up to $1$ TeV (and
larger than $2m_W$) can be discovered at LHC if the design luminosity is
achieved.

The most difficult range of the Higgs boson mass to probe at Hadron Colliders
is the so-called intermediate mass region, beyond the reach of LEP-200 and
below the mass, where the Higgs has a decay channel into two massive gauge
bosons ($WW, ZZ$), 90~GeV$<m_H<2 m_W(m_Z)$. Discovery of the intermediate mass
Higgs boson at Hadron Colliders is possible through the rare decays with light
leptons and photons in final states, $H\to~ZZ^*\to~l^+l^-l^+l^-$, $H\to\g\g$
\cite{wudka}, or associated $WH$ production  $WH\to l\nu\g\g$
\cite{kunszt}. The combination of these modes provides a possibility of
covering the whole intermediate mass region at LHC only in the case when the
$\g\g$ resolution is adequate to ~resolve the $H\to\g\g$ ~signal ~and high
~luminosity ($L = 10^5$~pb$^{-1}$) is achieved \cite{wudka,kunszt}.  For the SM
Higgs boson of intermediate mass primary decay mode is $H\to b\bar b$ but,
unfortunately, detection of Higgs through heavy quark pairs is impossible due
to large QCD backgrounds \cite{j1}.

The collision of high energy, high intensity photon beams at the Photon Linear
Collider (PLC), obtained via Compton backscattering of laser beams off linac
electron beams, provides another opportunity to search for an intermediate-mass
Higgs boson through the resonant production
\cite{gunion,borden,threejet,zerwas,brodsky,BBB} 
\begin{equation}
\g\g\to H\to b\bar b.
\end{equation}

Based on the $e^+e^-$ linear collider PLC will have almost the same energy and
luminosity, i.e. c.m. energy of 100--500 GeV and luminosity of the order of
$10^{33}$~cm$^{-2}$s$^{-1}$ \cite{PLC}.  Polarizing the linac electrons and
laser photons provides polarized backscatte\-red photons as well as photon
 energy
distribution needed.  Colliding like-handed electrons and photons results in a
flat distribution of backscattered photons and colliding oppositely handed
electrons and laser photons gives a peaked distribution of backscattered
photons with energy just below the $e^+e^-$-collider energy.

Extracting the intermediate mass Higgs signal in photon-photon collisions is a
hard task since a large number of $b\bar b/c\bar c$ background events must be
rejected \cite{gunion,borden,eboli,BBB}. The crucial assumption is that these
large backgrounds can  actively be suppressed by exploiting the polarization
dependence of the cross sections. Far above the threshold, the $\g\g\to q\bar
q$ cross section is dominated by initial photons in the $J_z=\pm 2$ helicity
state. Taking into account that the Higgs signal comes from the $J_z=0$
channel, polarized collisions can be used to enhance the signal  simultaneously
suppressing the background \cite{gunion,borden} (see also a detailed discussion
in \cite{threejet,BBB}).  The search for intermediate-mass Higgs requires not
only luminosity distribution dominating at $J_z=0$. Another important
requirement to $\g\g$-luminosity is that it must cover the entire intermediate
mass region.  Utilizing the broad photon-photon luminosity at fixed linac
energy of 125 GeV provides high $\g\g$-luminosity and high photon polarization
over the whole region of interest \cite{borden}.

To study the properties of the Higgs boson of known mass, the peaked
photon-photon luminosity spectrum is more convenient, it could be obtained by
choosing the collider energy so that the peak of the luminosity spectrum sits
at the Higgs boson mass.  The Higgs boson production in such collisions would
provide an accurate measurement of the $H\to\g\g$ coupling. This coupling is
induced at the one-loop level and receives contributions from all virtual
charged particles whose masses derive from the Higgs mechanism.  The
measurement of the $H\to\g\g$ coupling would give fundamental information about
the particle spectrum and mass generation mechanism of the theory.  In
addition, a new interesting method has been proposed \cite{grz,KKSZ} to measure
the parity of the Higgs states in linearly polarized photon-photon collisions.
It provides an opportunity to investigate nontrivial assignments of the quantum
numbers for Higgs particles in extended models such as supersymmetric theories
which include both scalar $0^{++}$ and pseudoscalar $0^{-+}$ states
\cite{SUSY}.

But the question remains how  QCD radiative corrections influence these
conclusi\-ons. Though it is known that far above the threshold the magnitude of
these corrections is moderate for unpolarized collisions \cite{KMS,DKZZ}, one
can expect that their effects will be especially large for the $q\bar q$
production in the $J_z=0$ helicity state, where the tree level contribution is
suppressed by the factor of $m_q^2/s$.  We presented our first results on the
one-loop QCD corrections to the $b\bar b/c\bar c$ quark pair production in
polarized photon-photon collisions in \cite{Berkeley}.  The lowest order cross
section, one-loop virtual corrections and gluon emission contributions were
shown to be of the same order of magnitude for the $b\bar b$ quark production
at $\sqrt{s_{\g\g}}\sim 100$~GeV in the $J_z=0$ channel, while QCD corrections
were found to be quite small for $J_z=\pm 2$.  The cross section of two-jet
final states in $b\bar b (g)$ production for $J_z=0$ even happened to be
negative for small values of $y_{cut}$ (which were used {\it e.g.} in
\cite{threejet}, where only radiative processes of $b\bar b g$ production were
taken into account for $m_b=0$). Here we present complete analytical results
for the one loop QCD corrections to the polarized cross sections and to the
matrix element of the process $\g\g\to b\bar b (c\bar c)$ retaining the full
dependence on the quark mass. We also carefully analyze the nature of the
peculiarities of the corrections for $b\bar b$ production at high energies in
the $J_z=0$ channel. We exploit a tensor reduction algorithm \cite{oldenb} to
express the cross sections and amplitudes in terms of the set of basic scalar
loop integrals. This leads to very compact expressions for one loop
contributions for polarized cross sections.

Recently, the next-to-leading order corrections to the heavy quark pair
production cross sections in polarized photon-photon collisions have been also
presented in ref. \cite{zaza}, where dimensional reduction have been used to
regularize both ultraviolet and infrared singularities.  Authors of ref.
\cite{zaza} considered the effects of the QCD corrections on the background
events for the Higgs signal from direct $b\bar b (g)$ production only for Higgs
boson production at the photon machine obtained from a 500~GeV $e^+e^-$ linear
collider. However, it has been explicitly shown in refs. \cite{eboli,BBB},
where all backgrounds from two-jet productions were included: direct, so-called
resolved and twice-resolved {\it etc.}, that the major contribution to the
background at 500~GeV is the 1-resolved coming from the gluonic content of the
photon, not the direct. Another backgrounds occur in processes where $c$-quarks
are produced instead of $b$-quarks. Since charm production is much larger than
$b$ production, due to the $c$-quark's stronger coupling to the photon, it
represents an important background even with good $b$-tagging and a low
probability that $c$-quark is misidentified as a $b$. The charm background is
also not considered in \cite{zaza}.

In present paper we study the effects of QCD corrections on the ability of PLC
to discover an intermediate mass Higgs boson for the case of a photon collider
operating at around the Higgs resonance based on the $e^+e^-$ collider with
energy $\sqrt{s}=250$~GeV. The resolved photon backgrounds are much less
significant at 250~GeV due to a steeply falling gluon spectrum.  We use
dimensional regularization to regulate ultraviolet divergencies and introduce a
small gluon mass $\lambda$ to regulate infrared divergencies. Analytical
results are given in Sections 2--4 and Appendices A, B.  In section 5 numerical
results are given. It is shown that taking account of the QCD corrections
reduces the signal-to-background ratio but the intermediate mass Higgs boson
signal is still expected with statistical significance of about 5$\sigma$.  The
influence of QCD corrections on the precise measurements of the Higgs
two-photon width using the peaked $\g\g$ luminosity is also considered.

\section{Born Cross Sections}
\setcounter{equation}{0}

At the tree level the cross sections of the quark-antiquark pair production in
the photon-photon fusion reaction
\begin{equation}
\g(p_1)\g(p_2)\to b(p_3)\bar b(p_4)
\end{equation}
for various helicity states of the colliding photons have the form
\begin{equation}
\frac{d\sigma^{Born}(J_z=0)}{dt} =\frac{12 \pi \alpha^2 Q_b^4}{s^2}
  \frac{m_b^2 s^2(s-2 m_b^2)}{t_1^2 u_1^2},
\end{equation}
and
\begin{equation}
\frac{d\sigma^{Born}(J_z=\pm2)}{dt} = \frac{12 \pi \alpha^2 Q_b^4}{s^2}
\frac{(t_1 u_1 - m_b^2 s)(u_1^2+t_1^2+2 m_b^2 s)}
{t_1^2 u_1^2}.
\end{equation}
We have introduced the following notation: $p_1$, $p_2$ are the photon momenta,
$p_3$ and $p_4$ are quark and anti-quark momenta.  $s=(p_1+p_2)^2$,
$t=(p_1-p_3)^2$, $u=(p_1-p_4)^2$, $t_1 = t-m_b^2$, $u_1 = u-m_b^2$.

The $J_z=0$ cross section is suppressed by $m_b^2/s$ factor at high energies
outside the very forward (backward) region.  However, the total cross section
integrated over the full phase space is not suppressed, so in the high energy
limit the $J_z=0$ cross section tends to a delta-function, peaking in the
forward (backward) direction.

\section{QCD One-Loop Corrections to the Matrix Element}
\setcounter{equation}{0}

Complete set of the diagrams describing $\alpha_s$-corrections for the process
$\gamma\gamma\to b\bar b$ is shown in Fig.~1.  This set contains genuine
one-loop diagrams and tree level diagrams involving counterterms.

The full one-loop matrix element (with the corresponding counterterms) can be
written  as follows
\begin{equation}
T^{(1)}(\gamma\gamma\to b \bar b) = 
e^2  g^2 a_i(s,t,u) \bar u(p_4) \hat O_i v(p_3)+
(t\leftrightarrow u, p_3\leftrightarrow p_4),
\end{equation}
where the set of the operators $\hat O_i$ is given by
\begin{center}
\begin{tabular}{l l}
$\hat O_1=(e_1 e_2)$,   &  $\hat O_6=\hat p_1 (e_1 e_2)$,\\
$\hat O_2=(e_1 p_3)(e_2 p_3)$, & $\hat O_7 = \hat p_1(e_1 p_3)(e_2 p_3)$,\\
$\hat O_3=\hat e_1 \hat p_1(e_2 p_3)-\hat e_2 \hat p_1 
(e_1 p_3)$,~~~~~~~~~~~~~  & $\hat O_8 =\hat e_1 \hat e_2$,\\
$\hat O_4=\hat e_1(e_2 p_3)$, & $\hat O_9 = \hat e_1 \hat e_2 \hat p_1$.\\
$\hat O_5=\hat e_2(e_1 p_3)$, &   ~~
\end{tabular}
\end{center}
Here $e_1$, $e_2$ are the photon polarization vectors.  To simplify the final
expressions, photon polarization vectors are chosen so as to fulfil the
relations $(e_2 p_1)=0$ and $(e_1 p_2)=0$.

The coefficients $a_i(s,t,u)$ expressed through the scalar one-loop integrals
defined in the Appendix~A are given in the Appendix~B. The algebraic
calculation of one-loop diagrams was carried out by using the symbolic
manipulation program FORM \cite{form}.  The finite parts of the counterterms
are fixed by the standard conditions of the on-mass shell renormalization
procedure.  The sum of the diagrams is UV-finite and gauge invariant. To avoid
infrared singularities, we have introduced an infinitesimal mass of the gluon
$\lambda$.

\section{QCD ~One-Loop ~Corrections to ~Polarized ~Cross Sections}
\setcounter{equation}{0}

The cross sections to order $\alpha^2 \alpha_s$ are determined by the
interference between one-loop and tree level contributions  given in the
previous sections. For the different helicities of two photons ($J_z=0$ and
$J_z=\pm2$) they have the form
\begin{eqnarray}
&&\frac{d\sigma(J_z=0)}{dt}  = \frac{\alpha^2 \alpha_s Q_b^4}{s^2} \nonumber\\
&&~~\frac{16 m_b^2 s}{t_1 u_1} \biggl\{
 (t+u)^2 D(s,t) + 2 m_b^2 C(s) +\frac{2 m_b^2 s}{t_1} C_1(t) \nonumber \\
&&~~ -\frac{3 t u +m_b^2 u +3 t^2 - m_b^2t -2 m_b^4}{t_1^2} B(t)
-\frac{2 (s^2-t_1 u_1)+s s_4}{2 t_1 u_1} \nonumber \\
&&~~ +\frac{s (u+t)}{2 t_1 u_1} \ln{ \bigl(\frac{\lambda^2}{m_b^2}\bigr)}+
(t\leftrightarrow u)\biggr\},
\end{eqnarray}

\begin{eqnarray}
&&\frac{d\sigma(J_z=\pm2)}{dt}  =
  \frac{2\ \alpha^2 \alpha_s Q_b^4}{s^2}\nonumber\\
&&~~\biggl\{\biggl(-\frac{8 m_b^2 s s_4}{t_1 u_1}+
\frac{8 s(s t_1 +8 m_b^4)}{u_1}-4(s u_1 -2 m_b^2 s-8m_b^4)\biggr)D(s,t)
\nonumber\\
&&~~+\frac{2 s (3 s^2 -2 t_1 u_1 -16 m_b^4}{t_1 u_1} C(s)
+\frac{2 s_4 (s^2+2 t_1 u_1 +2 m_b^2 s)}{t_1 u_1} C_1(s)
\nonumber\\
&&~~+\frac{8(2 m_b^2 (u-t) +t_1^2+s^2-8 m_b^2)}{u_1} C_1(t)\nonumber\\
&&~~-\frac{4Y}{t_1 u_1}\biggl(
\frac{t_1 t_2 u_1 -2 s(t^2+m_b^4)-8 m_b^2 t^2}
{t t_1^2 } B(t)-\frac{2 s}{s_4} B(s)
\nonumber\\
&&~~+\frac{u_1^2+t_1^2+2 m_b^2 s}{t_1 u_1}
\ln{\bigl(\frac{\lambda^2}{m_b^2}\bigr)}\biggr)
\nonumber\\
&&~~-\frac{32 m_b^2}{s u_1}(Y+u_1 u_2)-\frac{4}{u_1}
\biggl(\frac{t_1(3 u-m_b^2)}{u_1}-\frac{2m_b^4}{t}\biggr)+
\frac{4 m_b^2}{u}\nonumber\\
&&~~+(t\leftrightarrow u)\biggr\},
\end{eqnarray}
where $s_4=s-4 m_b^2$, $t_1 = t-m_b^2$, $u_1 = u-m_b^2$, $t_2=t+m_b^2$,
$u_2=u+m_b^2$ and $Y=ut-m_b^4$.

The explicit forms of the scalar functions $D$, $C$ and $B$ are given in the
Appendix A.  The IR divergencies are canceled out if one takes into account the
contribution of the soft gluon emission
\begin{eqnarray}
\frac{d\sigma^{soft}}{dt} = \frac{d\sigma^{tree}}{dt} R^{soft},
\end{eqnarray}
where
\begin{eqnarray}
R^{soft} &=& 
\frac{8\alpha_s}{3\pi}\Biggl\{
\left(-1+\frac{1}{\beta}(1-\frac{2 m_b^2}{s})
\ln\biggl(\frac{1+\beta}{1-\beta}\biggr)\right)
\ln\biggl(\frac{2k_c}{\lambda}\biggr)+\nonumber \\
&&\frac{1}{2\beta}\ln\biggl(\frac{1+\beta}{1-\beta}\biggr)+
\frac{1}{2\beta}\biggl(1-\frac{2 m_b^2}{s}\biggr)
\Biggl[Sp\biggl(\frac{-2\beta}{1-\beta}\biggr)
-Sp\biggl(\frac{2\beta}{1+\beta}\biggr)\Biggr]\Biggr\}
\end{eqnarray}
and $k_c$ is the soft gluon energy cut and $\beta=\sqrt{1-4m_b^2/s}$.

The cross section of hard gluon emission is evaluated numerically.  The matrix
element squared is calculated by using FORM and also by means of COMPHEP system
\cite{comphep}.  The results are quite lengthy to be presented here and
coincide with each other numerically.  The integration over three-particle
phase space is done by using Monte-Carlo integration routine VEGAS
\cite{lepage}. Special care is taken to handle sharp peaks of the cross section
arising when gluon is soft and when gluon is emitted along the quark or
anti-quark momentum and corresponding to the infrared and collinear
singularities. In order that VEGAS algorithm converged fast enough all the
singularities must be running along the axes of the integration variables (see
detailed discussion in \cite{FINST}). We take gluon energy, the denominator of
the quark propagator and quark and anti-quark production angles as integration
variables, so that the infrared and collinear as well as $t$-channel
singularities all run along some axis.

The total cross section for $b\bar b$ production can be cast into the form
\cite{KMS}
\begin{equation}
\sigma^{\gamma\gamma\to b\bar b(g)}_{\lambda_1\lambda_2}\,
=\,\frac{\alpha^2\,Q_b^4\,N_c}{s}
\,\,\Biggl[f_{\lambda_1\lambda_2}^{(0)}\,+\,\frac{4}{3}\frac{\alpha_s}{\pi}
\,f_{\lambda_1\lambda_2}^{(1)}\Biggr],
\end{equation}
where $f^{(0,1)}_{++,+-}$ depend on the dimensionless variable $s/(4m_q^2)$
only.  In  Table~1 the values of the functions $f^{(0,1)}_{++,+-}$ are
presented for various helicity states of initial photons and for the case of
unpolarized photon collisions versus $s/(4m_b^2)$.  Both $f^{(0)}_{++}$ and
$f^{(1)}_{++}$ are not suppressed at high energies, because no angular cut is
imposed in Table~1 ({\it cf.} Section~2). While $f^{(1)}_{+-}$ is always
positive and monotonically rising, $f^{(1)}_{++}$ has a minimum near to
$s/(4m_b^2)\sim 9$, where it is negative. The value of $f^{(1)}_{++}$ at
threshold is not zero due to a familiar Sommerfelds rescattering correction.
The values of one-loop correction function $f^{(1)}_{unpol}$ obtained for
unpolarized photon collisions
\begin{equation} 
f^{(0,1)}_{unpol} = 
\frac{1}{2}\left(f^{(0,1)}_{++}+f^{(0,1)}_{+-}\right) 
\end{equation} 
agree with the results of papers \cite{KMS,DKZZ} to the accuracy better than
$0.3\%$.

\begin{table}
\begin{center}
\begin{tabular}{|c|c|c|c|c|c|c|}\hline
$\frac{s}{4 m_b^2}$ & 
$f^{(0)}_{++}$ & $f^{(0)}_{+-}$ & $f^{(0)}_{unpol}$ & 
$f^{(1)}_{++}$ & $f^{(1)}_{+-}$ & $f^{(1)}_{unpol}$ \\ 
\hline
   1 &   0   &   0  &  0   &  124    &        0  &    62.0 \\
   4 &  26.3 & 27.2 & 26.7 &    7.05 &       82.2&    44.6 \\  
   9 &  27.0 & 44.5 & 35.8 & $-2.95$ &      129  &    62.8 \\   
  16 &  26.7 & 57.4 & 42.1 &   29.1  &      169  &    99.2 \\   
  25 &  26.4 & 67.7 & 47.1 &   72.6  &      208  &   140   \\ 
 100 &  25.6 & 101  & 63.3 &  292    &      388  &   340   \\ 
 400 &  25.3 & 135  & 80.3 &  618    &      724  &   671   \\ 
2500 &  25.2 & 181  & 103  & 1200    &     1530  &  1370    \\ \hline
\end{tabular}
\end{center}
\caption{Functions $f^{(0,1)}$ for various values of $\frac{s}{4 m_b^2}$ for
polarized and unpolarized incoming photons.}
\end{table}

The total cross sections calculated up to the order $\alpha^2\alpha_s$ are
given by the sum  of the tree-level contribution (Section 2), the interference
term  between the one-loop and tree-level contributions, and the tree level
contribution from  the quark pair production accompanied by the gluon emission
$\gamma\gamma\to~q\bar qg$.  The first two contributions lead to two parton
final states converting mainly into two jets, while the third one leads to the
three parton production converting both into two- and three-jet final states.
The reason is that three parton final states with collinear and/or soft gluon
will appear experimentally as two jets.  Moreover, only the sum of cross
sections of $q\bar q$ and $q\bar q g$ production with the soft or collinear
gluon is free from infrared divergencies and has no mass singularities in the
limit $m_q\to 0$.  So, as usual, we consider the three parton state to
represent the two-jet final state if the invariant mass of two partons is
sufficiently small
\begin{equation} 
s_{ij} < y_{cut} s_{\g\g},
\label{ycut}
\end{equation} 
where $s_{ij}=(p_i+p_j)^2$ is the invariant mass squared of two partons $i$ and
$j$ and $\sqrt{s_{\g\g}}$ is the total c.m.s.  energy of two colliding photons.

Fig.~2 shows the total ({\it i.e.} two-jet plus three-jet) and two-jet
($y_{cut}=0.08$) cross sections for the $b\bar b/c\bar c$ pair production in
polarized monochromatic $\g\g$ collisions.  Throughout we use a two-loop
expression for $\alpha_s(Q^2)$ with $\Lambda=200$~MeV, $Q^2=s$ and $N_F=5$ as
the number of flavors. We take $m_b=5$~GeV and $m_c=1.5$~GeV. The angular cut
$|\cos\theta|<0.7$ means a cut on the scattering angles of both quark and
anti-quark.  This choice is different from that used in \cite{zaza}, where only
the quark scattering angle is restricted.

While the QCD corrections for the $J_z=\pm 2$ photon helicities are quite
small, those for $J_z=0$ enhance $c\bar c$ production by an order of magnitude
or even larger. For the $b\bar b$ production the situation is more complicated:
the corrected total cross section is smaller than the tree level $\g\g\to b\bar
b$ cross section for $\sqrt{s_{\g\g}}< 85$~GeV and larger for larger energies.
The effect is more pronounced for the two-jet production.

For small values of $y_{cut}<0.04$ the two-jet differential cross section in
$J_z=0$ channel is even negative in some regions of the phase space. Fig.~3
gives the differential cross sections for the $b\bar b$ pair production versus
the scattering angle at $\sqrt{s_{\g\g}}=100$~GeV for various helicity states
of initial photons and various cuts.  As it is shown in Fig.~3a, for $J_z=0$
and $y_{cut}=0.04$ differential cross  section becomes negative in the central
region of the scattering angles.  This means that for the $b\bar b$ production
at $\sqrt{s_{\g\g}}\sim 100$~GeV all the three contributions (Born, virtual and
real gluon emission) are of the same order of magnitude and perturbation
expansion is not valid for too small values of $y_{cut}\leq 0.04$. This is
unlike the case of the $c\bar c$ production, where the real gluon emission
contribution still dominates for $y_{cut}=0.04$.

To elucidate the breakdown of the perturbative expansion for $b\bar b$
production in the $J_z=0$ channel at high energies we present the sum of
virtual and soft gluon emission contributions in the limit $s,-t,-u\gg m_b^2$
under the double logarithmic approximation
\begin{eqnarray}
\frac{d\sigma^{soft+virt}(J_z=0)}{d\cos\theta}
& =& -\, \frac{4\,\alpha_s\,\alpha^2\,Q_b^4}{s} 
\frac{m_b^2}{s} \left\{
-\, 2\,(\frac{s}{t}+\frac{s}{u})\ln^2(\frac{s}{m_b^2}) 
\right.
\nonumber \\
&&+ \,\frac{s^2}{t^2}\ln(\frac{s}{m_b^2})
\left(\ln(\frac{s}{m_b^2})+4\ln(-\frac{t}{m_b^2})\right) \nonumber \\
&&+ \,\frac{s^2}{u^2}\ln(\frac{s}{m_b^2})
\left(\ln(\frac{s}{m_b^2})+4\ln(-\frac{u}{m_b^2})\right)  \label{asympt} \\
&& + \,4\,\frac{s^2}{tu}\ln(\frac{s}{m_b^2})
\left(\ln(-\frac{t}{m_b^2})+\ln(-\frac{u}{m_b^2})\right)
\nonumber\\
&&\left.-\,2\,(\,\frac{s^2}{t^2}\,+\,\frac{s^2}{u^2}\,-\,2\,\frac{s}{t}
\,-\,2\,\frac{s}{u})\ln(\frac{s}{m_b^2})\,\ln(\frac{4k_c^2}{m_b^2})
	\right\}.\nonumber
\end{eqnarray}
It is this large negative double logarithmic contribution which makes two-jet
cross section negative.  Normally, (and this is the case for $J_z=2$ channel)
$\ln^2(s/m^2)$ terms are cancelled out. This is a consequence of the
Kinoshita-Lee-Nauenberg theorem \cite{KLN} stating that cross sections
integrated over all degenerate in energy final states are free from mass
singularities. However, this theorem is trivially fulfilled for $J_z=0$
channel, as Born cross section, virtual correction and soft gluon emission are
all equal to zero for $m_b=0$. The cross section of the hard gluon emission
$\sigma(\g\g\to b\bar b g)$ in principle is not suppressed in the $J_z=0$
channel in the limit $m_b=0$. But selecting only two-jet topologies for
$y_{cut}\ll 1$ we suppress also the cross section of the hard gluon emission.
It is worth mentioning, that cross section of $b\bar b g$ production also
contains subleading double logarithmic terms of the order ${\cal
O}\left(\alpha^2\alpha_sm_b^2/s\ln(s/m_b^2)\ln(s/k_c^2)\right)$, but they do
not completely cancel the double logarithmic terms (\ref{asympt}). Consider two
final state jets to lie in the central region of the detector with
$|\cos\theta|<\Delta\,z\ll 1$. Then $J_z=0$ cross sections are given by
\begin{eqnarray}
\Delta\sigma^{Born}&=&
\frac{192\pi \alpha^2 Q_b^4}{s}\frac{m_b^2}{s}\,\Delta z\nonumber\\
\Delta\sigma^{soft+virt}&=&
-\frac{384 \alpha_s \alpha^2 Q_b^4}{s}\frac{m_b^2}{s}\ln^2\frac{s}{m_b^2}
\,\Delta z \\
\Delta\sigma^{hard}&=&
\frac{128 \alpha_s \alpha^2 Q_b^4}{s}
\left(2y_{cut}\ln\frac{1}{2y_{cut}}-y_{cut}\right)\,\Delta z.
\nonumber
\end{eqnarray}
The last cross section was calculated in ref. \cite{threejet}.  We take
$k_c=\sqrt{s}/2$ in (\ref{asympt}) to take into account a partial cancellation
of double logarithmic terms between contribution (\ref{asympt}) and the hard
gluon emission contribution, calculated for finite value of $m_b\neq 0$. The
ratio of $\sigma^{soft+virt}/\sigma^{Born}$ is equal to
$-2\alpha_s/\pi\ln^2(s/m_b^2)$ and at 100~GeV virtual correction is $(-2.7)$
times larger than Born cross section of $b\bar b$ pair production! For a
$y_{cut}$ of 0.02, as it has been used in \cite{threejet}, virtual correction
is also $(-2.5)$ times larger than the cross section of $b\bar b g$ production
at $\sqrt{s_{\g\g}}=100$~GeV.  For $c\bar c$ production both virtual correction
and Born cross section are an order of magnitude smaller and total cross
section is dominated by $c\bar c g$ production contribution.  Therefore, the
approach of \cite{threejet}, where only contributions from the radiative
processes $\g\g\to c\bar c g,\ b\bar b g$ have been taken into account in the
limit $m_c=m_b=0$ and stringent value of $y_{cut}=0.02$ has been used to
select two-jet-like events, might be relevant for the $c\bar c$ production, but
is definitely not applicable for the $b\bar b$ production, where higher order
resummation of double logarithmic terms is necessary. However, for a loose
value of $y_{cut}=0.08$ and a nonzero value of the $b$-quark mass the cross
section of the radiative process $\g\g\to b\bar b g$ is large enough for the
total cross section to be always positive.

Note also that in the limit $m_b=0$ one-loop amplitude itself is infrared
finite, so we can calculate the $\alpha_s^2\alpha^2$ correction in the zero
quark mass limit by just squaring the one-loop amplitude
\begin{equation}
\frac{d\sigma^{(2)}(J_z=0)}{dt} = \, \frac{8\,\alpha_s^2\,\alpha^2\,Q_b^4}
{3\pi s^2}\,\frac{(t-u)^2}{tu}.
\label{square}
\end{equation}
However, numerically the contribution (\ref{square}) is negligibly small in the
central region.

As one can see from (4.1)--(4.4) the sum of the cross
sections of the $q\bar q$ production and $q\bar qg$ production with the soft
gluon does not depend on the gluon mass $\lambda$. The cancellation of the
dependence on  $k_c$ (the soft gluon energy cut) was checked numerically in
each case of the two- and three-jet event selection cuts used . For testing the
integration method the 2-jet cross section was calculated in two different
ways: as a difference between total and 3-jet cross sections and directly using
the corresponding cuts for the invariant masses (\ref{ycut}). In both cases
similar values for the 2-jet cross section were obtained.

For comparison with the results of paper \cite{zaza} we have calculated the
total (2+3 jet), 3-jet and 2-jet cross sections using their cuts.  The results
are slightly different. We obtained that the total cross sections calculated
using our expressions are about $15-25\%$ larger than in \cite{zaza} in both
cases of $J_z=0,\pm2$.  Our 3-jet cross section is about $5\%$ larger for
$J_z=\pm2$ and about $10-15\%$ smaller for $J_z=0$ than those from \cite{zaza}.
But these small differences lead to  larger discrepancies for the 2-jet cross
sections. For example, at $\sqrt{s}=40$~GeV for $J_z=0$ we have got 2-jet cross
section which is about $80\%$ larger than that from \cite{zaza}.

\section{Higgs ~Boson ~Production at ~Photon ~Linear Collider}
\setcounter{equation}{0}

As has been mentioned above, a photon linear collider provides an excellent
possibility of searching for the intermediate mass Higgs boson through the
resonant $\g\g\to H\to b\bar b$ production. In this region of mass, the
dominant background to such a process will be the continuum production of heavy
quark pairs.  As it is discussed previously, at the tree level the quark pair
production cross section far above the threshold is suppressed by a factor of
$m_q^2/s$ if two initial photons are in the $J_z=0$ helicity state from which
the Higgs signal comes.  So, the use of the $J_z=0$ dominated photon-photon
luminosity distribution reduces the number of background events.  Another
requirement is that the $\g\g$-luminosity must cover the entire intermediate
mass region \cite{borden}.  We make here the same assumptions as in
\cite{borden}, {\it i.e.}, we choose the broad photon-photon luminosity
spectrum resulting from polarized linac electrons and laser light for
$\lambda_{\g}\lambda_e>0$, $\lambda_e=0.9$, $\lambda_{\gamma}=1$, parameter
$x=4.8$ and geometric factor $\rho = 0.6$ \cite{borden,telnov}. We also assume
that the linac beam energy equals  125~GeV and the integrated effective
luminosity is 20~fb$^{-1}$.  In Fig.~4, the luminosity distributions for the
machine parameters mentioned are plotted.

Our task is to compare the signal and background event rates taking into
account the QCD corrections.  Fig.~5 shows the event rates of signal and
background two-jet final states in photon-photon collisions at tree level (a)
and taking into account QCD corrections (b).  We ignore here the backgrounds
from the $e\g\to~ eZ\to~ eb\bar b$ and $\g\g\to~f\bar fZ$ processes
\cite{HZ}, which are essential for $m_H\sim m_Z$. The backgrounds coming
from the resolved photon contributions $\g g\to b\bar b,\ c\bar c$ are also
shown. While resolved photon contributions make it very hard to observe the
intermediate mass Higgs signal at the 500~GeV linear collider \cite{eboli}
(see, however, recent analysis \cite{BBB}, where conclusion is done that using
optimized cuts still it will be possible to extract Higgs signal in the range
110--140~GeV at 500~GeV), these backgrounds are much less significant at
250~GeV due to a steeply falling gluon spectrum (see also \cite{borden,KKSZ}).
QCD corrections to the Higgs decay into $b\bar b$ \cite{Hbb} are also taken
into account. We use a cutoff $|\cos\theta|<0.7$ in the laboratory frame and
not in the c.m.s. frame as in \cite{borden}. Cut in the laboratory frame gives
a slightly better statistical significance of the Higgs signal. Finally, we
assumed 5\%  $c\bar c$-to-$b\bar b$ misidentification probability. Thus, the
combined background ({\it i.e.} $b\bar b + 0.05 c\bar c$) is represented by the
dotted line and can be compared with the signal denoted by the solid line
(Fig.~5b).

Fig.~6 presents the statistical significance of the Higgs boson signal
estimated from the tree level and the one-loop calculations including the
resolved photon contributions. This plot assumes  a 50\% $b\bar b$-tagging
efficiency for the $b\bar b$ final states and the resolution for reconstructing
the invariant mass of  two-jet events to be Gaussian with $FWHM=0.1m_H$. From
this figure one can conclude that it is advantageous to select two-jet final
states and to impose the angular cut in the laboratory frame. The account of
the QCD corrections reduces the statistical significance of the Higgs signal
almost by a factor of two in comparison with the tree-level result.
Nevertheless, the intermediate mass Higgs boson can be observed in $\g\g$
collisions at least at the level of 5$\sigma$ in the mass interval from 80 to
160~GeV.

On the other hand, if the intermediate mass Higgs boson is discovered, the PLC
gives the best opportunity for measuring the decay width of the Higgs boson
into two photons by measuring the resonant production rate of the  $b\bar b$
pairs, which is proportional to the $H\to\g\g$ decay width and the $H\to~b\bar
b$ decay branching ratio.  To measure the two  photon width it is more
convenient to use the  peaked $\g\g$ luminosity distribution, which is obtained
for the following combination of parameters: $\lambda_e\lambda_\g<0, \rho>1$.
For this configuration, the luminosity distribution is  fairly monochromatic
($\sim10\%$  energy spread) and  very highly polarized ($>95\%$) \cite{borden}.
Fig.~7 shows  the  corresponding $\g\g$ luminosity for both helicity states  of
two photons. The effective luminosity of 20~fb$^{-1}$ is assumed and parameter
$\rho$ is taken to be equal to $3.0$ to suppress the low invariant mass tail of
the luminosity distribution function.

Considering only $b\bar b$-final states,  Fig.~8 gives the expected event rates
for the signal and background processes at the  tree level and including
$\alpha_s$-corrections. Collider energy is chosen so that the peak of the
luminosity spectrum coincides with the mass of the Higgs boson. As in
\cite{borden}, it is assumed that a window in   the invariant mass of
$\pm2\sigma$ around the Higgs mass is used for the measurement  and the
resolution for reconstructing the invariant mass of  two-jet events is Gaussian
with FWHM=0.1$M_H$.  In both cases the angular cut $|\cos\theta|<0.7$ in the
laboratory frame is used, as analogous cut in the c.m.s. frame gives smaller
signal to background ratio.

In  Fig.~9  the expected statistical errors in the measured two-photon width of
the Higgs boson are plotted. For comparison we have presented tree level
results with the cut $|\cos\theta|<0.7$ in both c.m.s. and laboratory frames,
and one-loop QCD corrected results in the laboratory frame.  Again,
20~fb$^{-1}$ of the integrated effective luminosity and 50\% $b\bar b$ tagging
efficiency with the 5$\%$ $c\bar c$ contamination are assumed.  As one can see
from Fig.~9, for the above mentioned parameters the two-photon Higgs boson
width can be measured with the statistical error of 6-9$\%$ in the wide
range of Higgs mass  $40\div 150$~GeV.  Of course, a more detailed analysis,
including full detector simulation, could somewhat modify our estimates  of the
influence of QCD radiative corrections on the statistical significance of the
Higgs signal in photon-photon collisions.

\section{Conclusions}
\setcounter{equation}{0}

In the present paper we consider the influence of the QCD correcti\-ons on the
background rates for the intermediate mass Higgs boson signal in the process
$\g\g\to H \to b\bar b$. We have derived compact analytical expressions for the
$\alpha_s$-corrections to the matrix element of the process $\g\g\to b\bar
b(c\bar c)$ and have calculated QCD corrected heavy quark pair production cross
sections in polarized $\g\g$ collisions. The total cross sections to the order
$\alpha^2\alpha_s$ are calculated retaining the dependance on the quark mass.
The cross sections are given by the sum of the tree level cross section,
contribution of the interference term between the $\alpha_s$-correction  and
the Born amplitude, and the cross sections of quark pair production accompanied
by the real gluon emission. The last contribution generates 3-jet events and
2-jet events in the case of radiation of soft or collinear gluon. The
contribution of QCD radiative corrections is crucial in the $J_z=0$ channel for
$c\bar c$ production enhancing the $c\bar c$ event rate by an order of
magnitude due to a real gluon emission contribution, which is not suppressed by
a factor of $m_q^2/s$ as is the Born cross section. For $b\bar b$ production
the situation is more subtle. For small values of $y_{cut}\leq 0.04$, used to
separate two- and three-jet final state topologies, virtual correction is
negative and of the same order of magnitude as Born and $b\bar b g$ production
contributions, so that two-jet cross section in the central region is even
negative. The origin of this large negative contribution lies in the double
logarithmic contribution ${\cal
O}\left(\alpha^2\alpha_sm_b^2/s\ln^2(s/m_b^2)\right)$. Thus, for loose values
of $y_{cut}=0.08$ backgrounds coming from continuum $b\bar b (g)$ production
are negligible in comparison to ones from $c\bar c (g)$ production and one
might speculate that the same situation takes place for more stringent values
of $y_{cut}$ \cite{threejet}. However, strictly speaking, to safely estimate
$b\bar b (g)$ background for small values of $y_{cut}$ the summation of double
logarithmic terms is needed.

As a result, we have calculated the backgrounds for the Higgs signal in two-jet
events at PLC. It has been shown that the intermediate mass Higgs boson can be
observed at the level of 5$\sigma$ in the mass interval from 40 to 150 GeV for
broad spectrum of photon-photon luminosity distribution using suitable cuts.
We have also considered the influence of $\alpha_s$-corrections on the
measurement of the two-photon width of the Higgs boson using the peaked $\g\g$
luminosity.  It is shown that for the integrated luminosity of 20~fb$^{-1}$ a
measurement of the intermediate mass Higgs boson width $\Gamma(H\to\g\g$) is
possible with the statistical error of about 6-9$\%$.

\vspace*{1cm}

We are grateful to D.~Borden and O.~\' Eboli for helpful discussions.  This
work was supported in part by the International Science Foundation grant NJR000
as well as in part by the joint ISF-RFBR grant NJR300, and in part by the INTAS
grant 93-1180.

\section*{Appendix A}
\setcounter{equation}{0}
\def\thesection{A}

The  scalar four- and three-point functions used are given by the expressions:
\begin{eqnarray}
&&D(s,t) = \frac{1}{i \pi^2}\nonumber\\
&&\int\frac{d^4q}
{(q^2-m_b^2)((q+p_1)^2-m_b^2)((q+p_1+p_2)^2-m_b^2)((q-p_4)^2-\lambda^2)}
\nonumber\\
&&~~= \frac{2}{\beta s (m_b^2-t)}\Biggl\{Sp\biggl(\frac{1}{\beta}\biggr)-
Sp\biggl(-\frac{1}{\beta}\biggr)\nonumber\\
&&~~+\ln\biggl(-\frac{1-\beta}{1+\beta}\biggr)
\biggl[\ln\biggl(1-\frac{t}{m_b^2}\biggr)-\frac{1}{2}
\ln\biggl(\frac{\lambda^2}{m_b^2}\biggr)\biggr]\Biggr\};
\end{eqnarray}
 where $p_1^2=0$, $p_2^2=0$, $p_3^2=m_b^2$ and $p_4^2=m_b^2$, 
$\beta=\sqrt{1-\frac{4 m_b^2}{s + i0}}$.

\begin{eqnarray}
C_1(s) &=& \frac{1}{i\pi^2}\int\frac{d^4q}
{(q^2-m_b^2)((q+p_1)^2-m_b^2)((q+p_1+p_2)^2-\lambda^2)}\nonumber\\
&=&\frac{1}{\beta s} \Biggl\{
\ln\biggl(\frac{\lambda^2}{m_b^2}\biggr
)\ln\biggl(-\frac{1-\beta}{1+\beta}\biggr)
\nonumber\\
&&+ Sp\biggl(\frac{2}{1+\beta}\biggr)-Sp\biggl(\frac{2}{1-\beta}\biggr)
-2\biggl[Sp\biggl(\frac{1}{\beta}\biggr)-Sp\biggl(-\frac{1}{\beta}
\biggr)\biggr]\Biggr\};
\end{eqnarray}
$p_1+p_2+p_3=0$, $p_1^2=s$, $p_2^2=m_b^2$ and $p_3^2=m_b^2$.

\begin{eqnarray}
C_1(t) &=& \frac{1}{i\pi^2}\int\frac{d^4q}
{(q^2-m_b^2)((q+p_1)^2-\lambda^2)((q+p_1+p_2)^2-m_b^2)}\nonumber\\
&=&\frac{1}{t-m_b^2}\biggl\{\frac{\pi^2}{6}-Sp\biggl(\frac{t}{m_b^2}
\biggr)\biggr\};
\end{eqnarray}
$p_1+p_2+p_3=0$, $p_1^2=t$, $p_2^2=m_b^2$ and $p_3^2=0$.

\begin{eqnarray}
C(s) &=& \frac{1}{i\pi^2}\int\frac{d^4q}
{(q^2-m_b^2)((q+p_1)^2-m_b^2)((q+p_1+p_2)^2-m_b^2)}\nonumber\\
&=& \frac{1}{2s}\ln^2\left(-\frac{1-\beta}{1+\beta}\right);
\end{eqnarray}
$p_1+p_2+p_3=0$, $p_1^2=0$, $p_2^2=0$ and $p_3^2=s$.

Only the following combinations of two-point functions are present in the final
answer
$$ B(s) = \bar B(s)-\bar B(m_b^2),~~~~~~~~~~~~~~B(t) = \bar B(t)-\bar
B(m_b^2),
$$
where $\bar B(s)$ and $\bar B(t)$ are the following scalar integrals:
\begin{equation}
\bar B(s) = \frac{1}{i\pi^2}\int\frac{d^4q}
{(q^2-m_b^2)((q+p)^2-m_b^2)},
\end{equation}
$p^2=s$, and
\begin{equation}
\bar B(t) = \frac{1}{i\pi^2}\int\frac{d^4q}
{(q^2-m_b^2)(q+p)^2},
\end{equation}
$p^2=t$.
Thus,
\begin{equation}
B(t) = -\biggl(1-\frac{m_b^2}{t}\biggr)\ln\biggl(1-\frac{t}{m_b^2}\biggr),
\end{equation}
and
\begin{equation}
B(s) = \beta \ln\biggl(-\frac{1-\beta}{1+\beta}\biggr).
\end{equation}

The functions $C_1(u)$ and $B(u)$ can be obtained  by replacing $t\to u$ in
$C_1(t)$ and $B(t)$.

\section*{Appendix B}
\setcounter{equation}{0}
\def\thesection{B}

\begin{eqnarray}
&&a_1(s,t,u) = \frac{m_b (u-t)}{Y^2}\bigl[s t_1 (7 t_1^2 +2 m_b^2 t_1-4 m_b^4)+
2 t_1^3(t_1-4 m_b^2)+4 s^2 t^2\bigr] D(s,t)\nonumber\\
&&~~ +\bigl(\frac{4 m_b^3}{Y}-\frac{m_b t_1 t_2 (u-t)}{Y^2}\bigr)
(s C(s)+(u-t)C_1(s)+2 t_1 C_1(t))+\frac{2 m_b s t}{Y} C(s)\nonumber\\
&&~~+\frac{4 m_b t t_1}{Y}C_1(t)+
2 m_b t_2 \bigl(\frac{1}{Y}-\frac{1}{t t_1}\bigr) B(t)
-\frac{2 m_b t_1}{Y} B(s)+2 m_b \bigl(\frac{m_b^2}{t t_1}
-\frac{4 m_b^2}{Y}\bigr);
\end{eqnarray}

\begin{eqnarray}
&& a_2(s,t,u) = \frac{4 m_b s}{Y^3}(m_b^4 u +t^3-2 m_b^4 t)
\bigl[(u-t)(t_1 D(s,t)-C_1(s))-s C(s)\nonumber\\
&&~~-2 t_1 C_1(t)\bigr]+\frac{8 m_b s t_1 t_2}{Y^2}(t_1 D(s,t)-C_1(s))+
\frac{4 m_b s (t (u-t)-t_1 t_2)}{t_1 Y^2} B(t)\nonumber\\
&&~~ +\frac{4 m_b}{Y}\biggl(\frac{2(m_b^2 u+3 m_b^2 t_1 -t^2)}{Y}-
\frac{2 t+s+6 m_b^2}{s_4}\biggr)B(s)\nonumber\\
&&~~+\frac{4(Y+t_1 u)}{m_b t_1 Y}+\frac{4 m_b(4 t+5 s)}{s_4 Y}+
\frac{32 m_b^3 s}{Y^2} -3;
\end{eqnarray}

\begin{eqnarray}
&&a_3(s,t,u) = \frac{2 m_b(t_1(t-u)-3 Y)}{Y^2}
\bigr[t t_1 D(s,t)+s C(s)-(t-u)C_1(s)\nonumber\\
&&~~ +2 t_1 C_1(t) \bigr]+2 m_b \biggl(6-\frac{t t_1}{Y}
 - \frac{m_b^2 (u-t)}{Y}\biggl(1+
\frac{m_b^2 s}{Y}\biggr)\biggr) D(s,t)\nonumber\\
&&~~+\frac{4 m_b t_1}{Y} C_1(s)+
\biggl(\frac{4 m_b}{Y}-\frac{2 m_b}{t t_1}\biggr) B(t)+
\frac{4 m_b (t-u)}{s_4 Y} B(s)\nonumber\\
&&~~+ \frac{16 m_b u_2}{s_4 Y}+\frac{2 m_b}{t t_1};
\end{eqnarray}

\begin{eqnarray}
&&a_4(s,t,u) = \biggl(\frac{11 m_b^2(2 t t_1+m_b^2 s)-7 t_1(m_b^4+t_1 s)+
3 m_b^4 u_1}{Y}-2 u\nonumber\\
&&~~+\frac{(u-t) t_1^3 t_2}{Y^2}\biggr) D(s,t)
+\frac{4 m_b^2 t_2-2 t(t+3 u)}{Y}C_1(s)+\nonumber\\
&&~~\biggl(\frac{2 u+5 t}{Y}-\frac{m_b^4 s+t t_1^2}{Y^2}\biggr)
\bigr[(u-t)(m_b^2 D(s,t)+ C_1(s)) +s C(s)\bigr]+\nonumber\\
&&~~\biggl(\frac{4(m_b^2 s+3 t t_1)}{Y}+\frac{2(u-t)t_1^3}{Y^2}
-\frac{4t}{t_1}\biggr)C_1(t)-2\biggl(\frac{t_1}{Y}+\frac{m_b^2(3 t-m_b^2)}
{t t_1^2}\biggr)B(t)\nonumber\\
&&~~+\frac{2 s t_2}{s_4 Y}B(s)-\frac{8 m_b^2 (t-u)}{s_4 Y}
+\frac{8m_b^2}{t_1^2} -\frac{2}{t};
\end{eqnarray}

\begin{eqnarray}
&&a_5(s,t,u) =\biggl(\frac{(t^3 t_2-6m_b^4 t t_1+ m_b^6(3s -t_2))(u-t)}
{Y^2}\nonumber\\
&&~~+\frac{m_b^2 t (5 u_1-6 m_b^2)+3 t^2 t_2+u t_1 t_2}{Y}\biggr)D(s,t)
\nonumber\\
&&~~-\frac{s(2 m_b^2 t_1^2-s(t_2^2+2 m_b^4))}{Y^2}C(s)
-\frac{2 s (2 m_b^2 t^2 (u-t)-(t^2-3 m_b^4) t_1 t_2)}{t_1 Y^2}C_1(t)+
\nonumber\\
&&~~\biggl(\frac{2(m_b^2(u-t)-t t_2)}{Y}
-\frac{m_b^4(4 m_b^2-3 s) (u-t)-t_1 t_2 (t_2^2-8 m_b^4)}{Y^2} +1\biggr)C_1(s)
\nonumber\\
&&~~+\biggl(\frac{6 t_2}{t_1^2}-\frac{2(t+3 m_b^2)}{Y}\biggr)B(t)+
\frac{2 s_4(t-3 m_b^2)-8 m_b^2 t_2}{s_4 Y}B(s)\nonumber\\
&&~~+\frac{8m_b^2(2 t_2+3 s_4)}{s_4 Y}-\frac{4 t_1+6 m_b^2}{t_1^2}+
\frac{4}{t_1}\ln{\biggl(\frac{\lambda^2}{m_b^2}\biggr)};
\end{eqnarray}

\begin{eqnarray}
&&a_6(s,t,u) = -\frac{t_1 t_2 (u-t)}{Y^2}\bigr[s_4 t_1 D(s,t)+(u+t)C(s)-
s_4 C_1(s)-2 t_2 C_1(t)\bigr]\nonumber\\
&&~~+2 m_b^2\biggl(4 D(s,t)-\frac{m_b^2 (u-t)^2}{Y^2} C(s)-\biggl(\frac{2 t_2}{Y}-
\frac{4}{t_1}\biggr) C_1(t)\biggr) \nonumber\\
&&~~-\biggl(\frac{2 s_4 t_1}{s Y}+\frac{8 m_b^2}{s t_1}-\frac{4}{t_1}-
\frac{16 m_b^2}{t_1^2}-\frac{2}{t}\biggr)B(t)+\frac{2 t_2}{Y}B(s)\nonumber\\
&&~~-\frac{8 m_b^2(t-u)}{s Y}-\frac{2}{t}
+\frac{16 m_b^2 (t_1-2 s)}{s t_1^2}
+\frac{4}{t_1}\ln{\biggl(\frac{\lambda^2}{m_b^2}\biggr)};
\end{eqnarray}

\begin{eqnarray}
&&a_7(s,t,u) = \frac{ 4(m_b^2 s (u_1-4 t_1)+t_1^2(u_1-s))}{Y^3}
\bigl[s_4 t_1^2 D(s,t) -s t_2 C(s)\nonumber\\
&&~~-s_4 t_1 C_1(s)
-2 t_1 t_2 C_1(t)\bigr] +\frac{8 m_b^2 s s_4(m_b^2 s-u_1 t_1)}{Y^3}
\bigl[t_1 D(s,t)-C_1(s)\bigr]\nonumber\\
&&~~+\frac{4(t_1^3 u_1-m_b^2 s (3 t+m_b^2)^2)}{t t_1 Y^2}(1-B(t))-
\frac{8 s t t_2 (2 m_b^2 -t_1)}{t t_1 Y^2} B(t)\nonumber\\
&&~~+\frac{4(2 t_1 t_2-3 Y+2 m_b^2 (u-t))}{Y^2} B(s);
\end{eqnarray}

\begin{eqnarray}
&&a_8(s,t,u) = -m_b\biggl(\frac{s(u-t)m_b^2}{Y}+3 u_1-t_1\biggr) D(s,t)+
\frac{m_b s u_1}{Y} C(s)\nonumber\\
&&~~+\biggl(4 m_b-\frac{m_b s u_2}{Y}\biggr) C_1(s) +\biggl(\frac{2 m_b^3 s}
{Y}-2 m_b\biggr) C_1(t) +\frac{m_b t_2}{t t_1} B(t)\nonumber\\
&&~~ +\frac{4 m_b}{s_4} B(s) -\frac{4 m_b}{t_1}+\frac{m_b}{t}-\frac{8
m_b}{s_4}; 
\end{eqnarray}

\begin{eqnarray}
&&a_9(s,t,u) = -\frac{s t_2 (t_2-s_4)}{Y} D(s,t) +\frac{s(t_2 -s_4)}{Y} C(s)
\nonumber\\
&&~~+\biggl(\frac{(m_b^2-s_4)(u-t)-t_1 t_2}{Y}+1\biggr) C_1(s)+
\biggl(\frac{2 t_1(t_2-s_4)}{Y}-\frac{4 m_b^2}{t_1}\biggr) C_1(t)\nonumber\\
&&~~-\frac{(3t+m_b^2) t_2}{t t_1^2} B(t) +\frac{16 m_b^2}{t_1^2}+\frac{1}{t}-
\frac{2}{t_1}\ln{\biggl(\frac{\lambda^2}{m_b^2}\biggr)}.
\end{eqnarray}

\newpage
\section*{Figures}
\parskip=\baselineskip

\begin{figure}[h]
\setlength{\unitlength}{1cm}
\begin{picture}(16,7.5)
\put(3,0){\epsfig{file=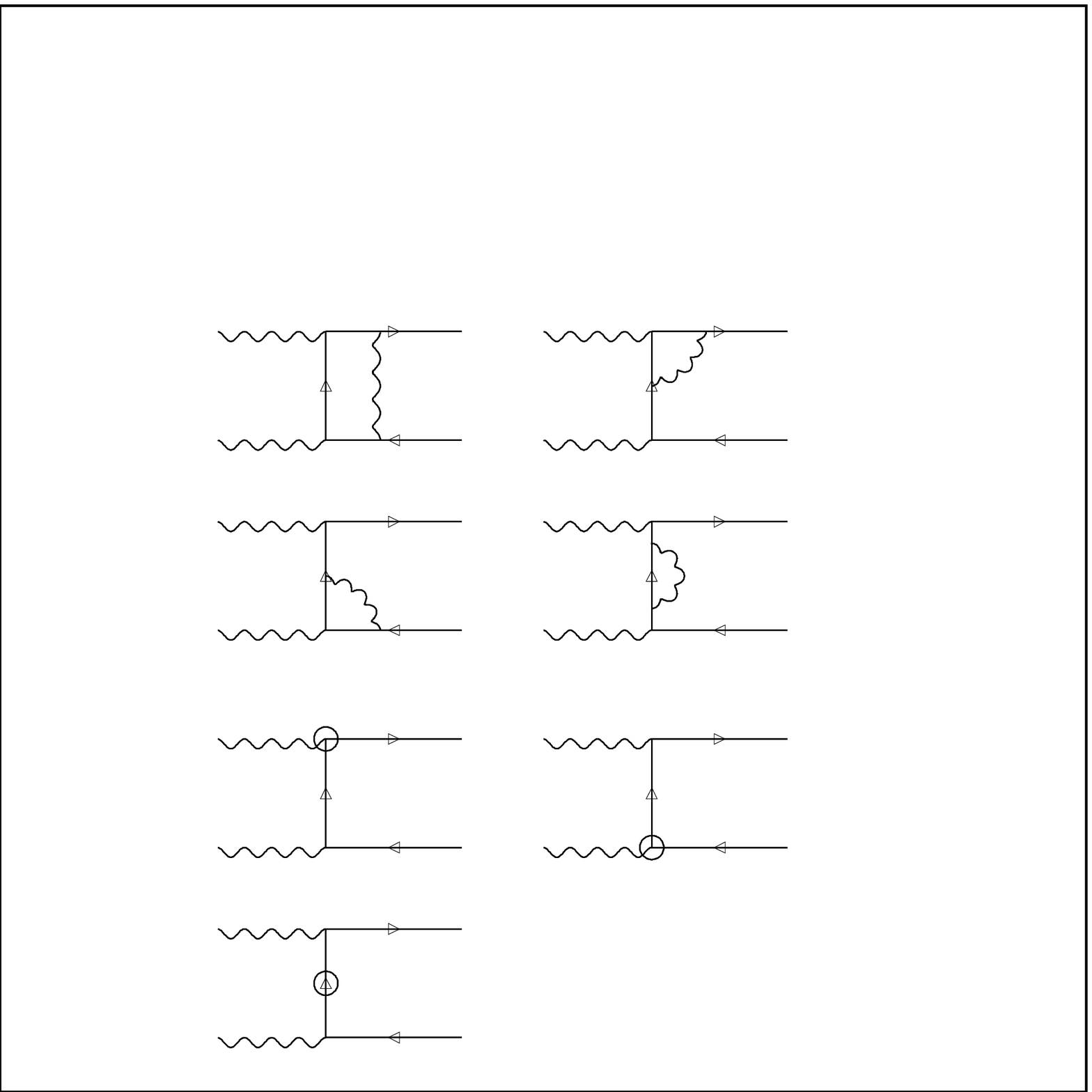,width=7.5cm}}
\end{picture}

\noindent{
Fig.~1. Diagrams for QCD ${\cal O}(\alpha^2\alpha_s)$-corrections to the heavy
quark production in $\g\g$ collisions.
}
\end{figure}

\begin{figure}[t]
\setlength{\unitlength}{1cm}
\begin{picture}(16,16)
\put(0,8){\epsfig{file=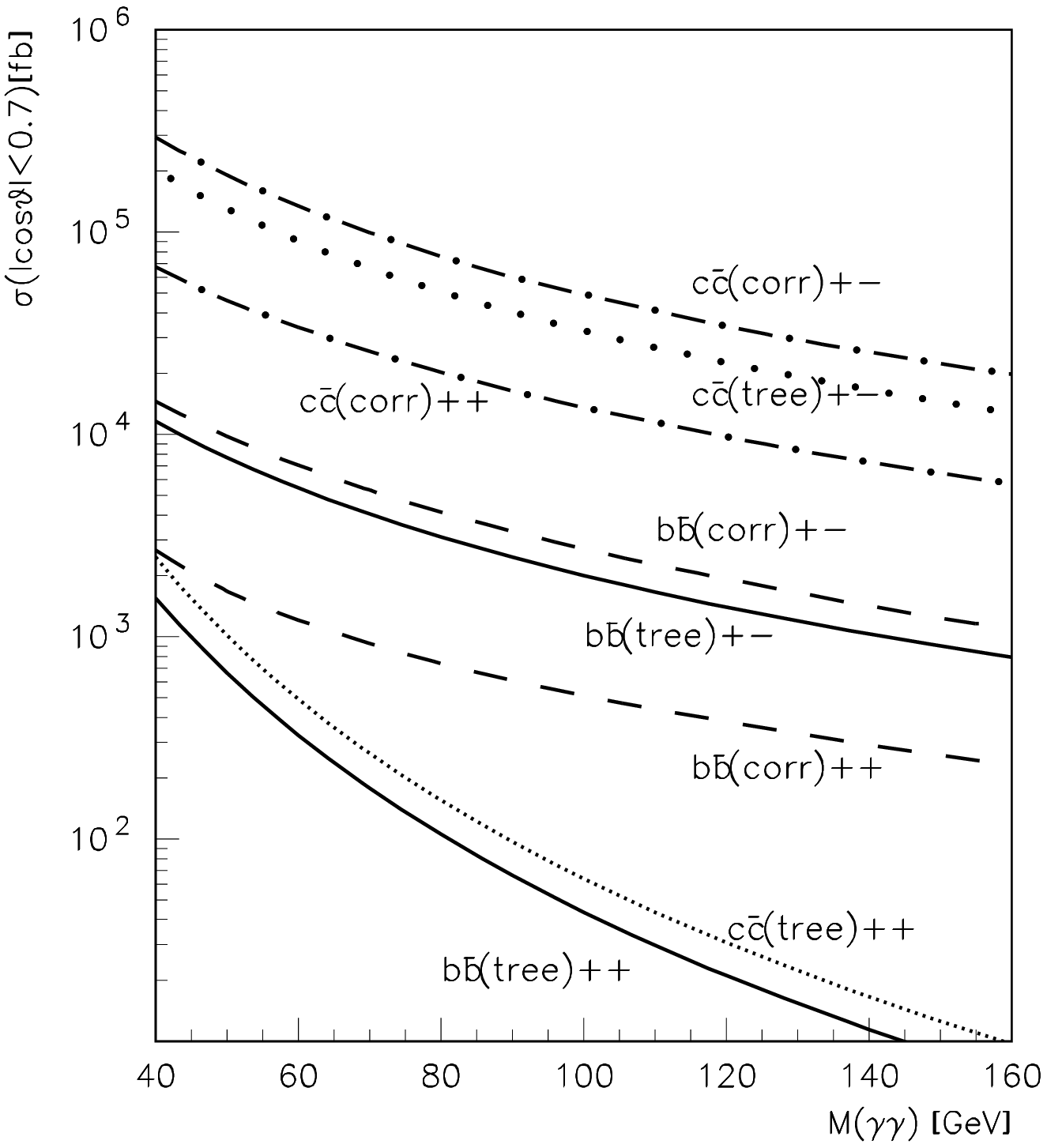,width=7.5cm}}
\put(8,8){\epsfig{file=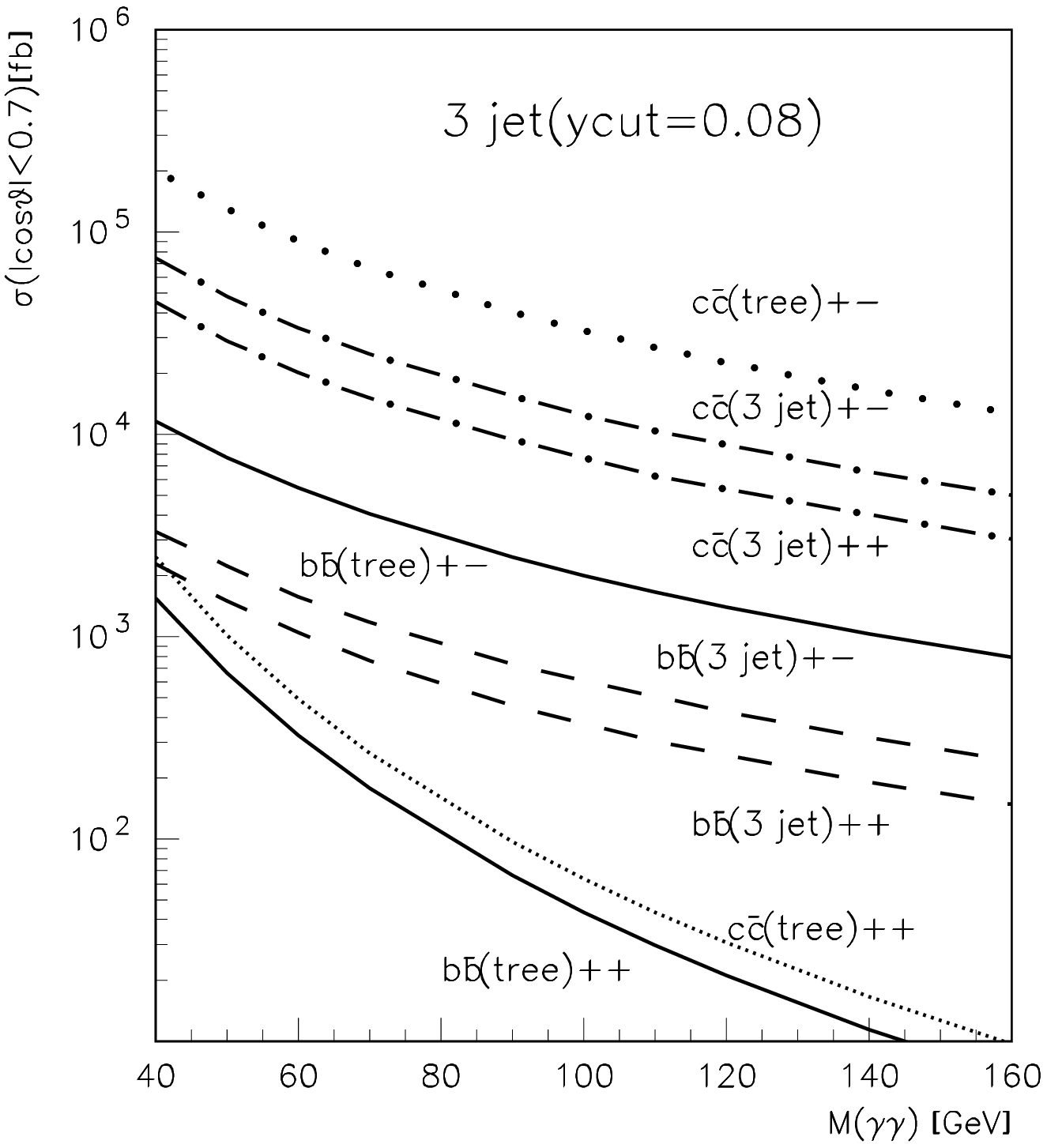,width=7.5cm}}
\put(3,0){\epsfig{file=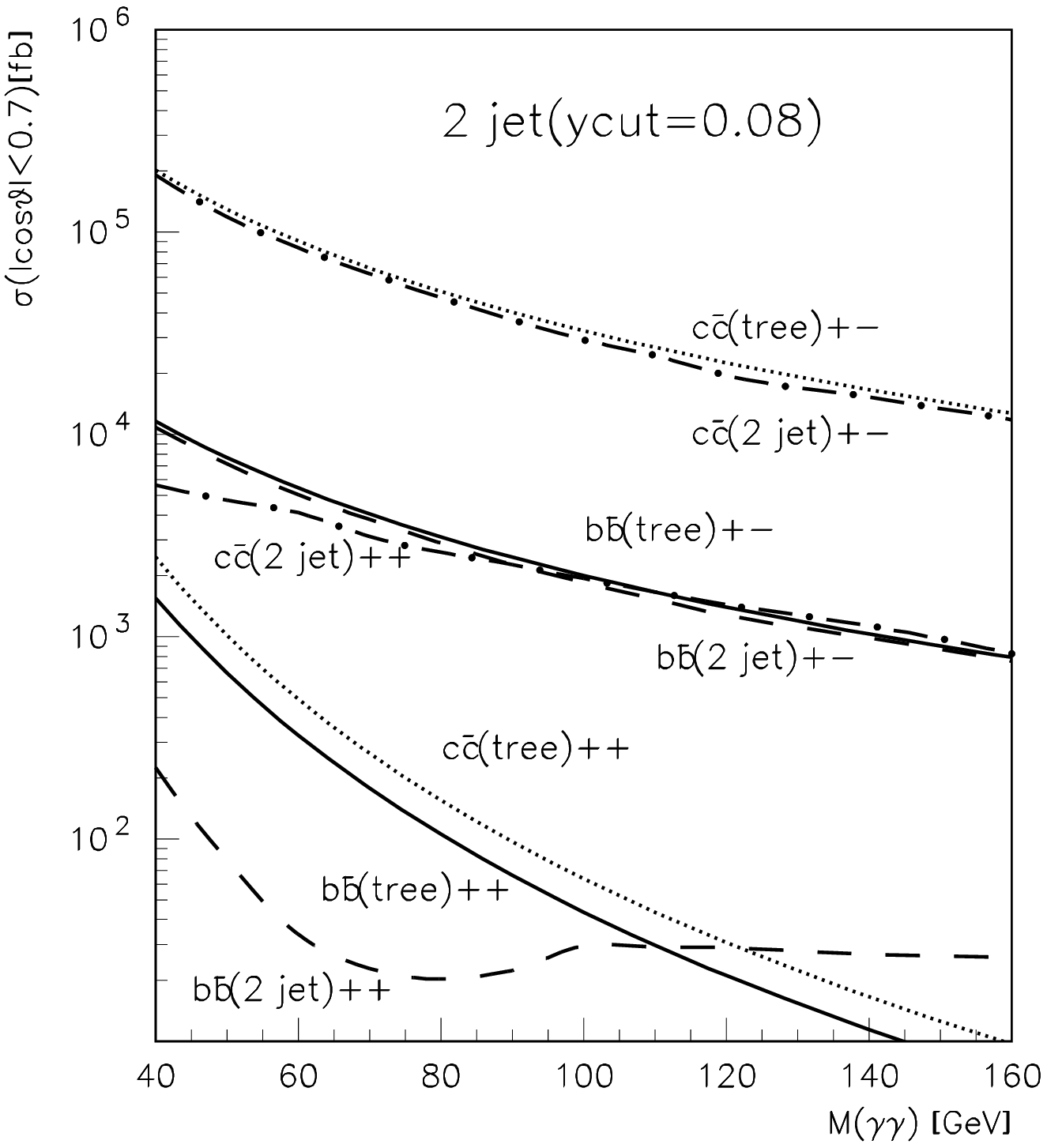,width=7.5cm}}
\end{picture}

\noindent{
Fig.~2. $b\bar b/c\bar c$-pair production cross sections in $\g\g$ collisions
for various helicity states of colliding photons. Solid lines correspond to
$\g\g\to~b\bar b$ cross sections at tree level, dashed lines -- to the
radiatively corrected $b\bar b$ production cross sections; dotted lines -- to
$\g\g\to~c\bar c$ cross sections at tree level and dash-dotted lines -- to the
corrected  $c\bar c$-pair productions; (a) total cross sections (i.e.  two-jet
plus three-jet),  (b) and (c) three- and two-jet cross sections, respectively,
with $y_{cut}$=0.08.
}
\end{figure}

\begin{figure}[t]
\setlength{\unitlength}{1cm}
\begin{picture}(16,7.5)
\put(0,0){\epsfig{file=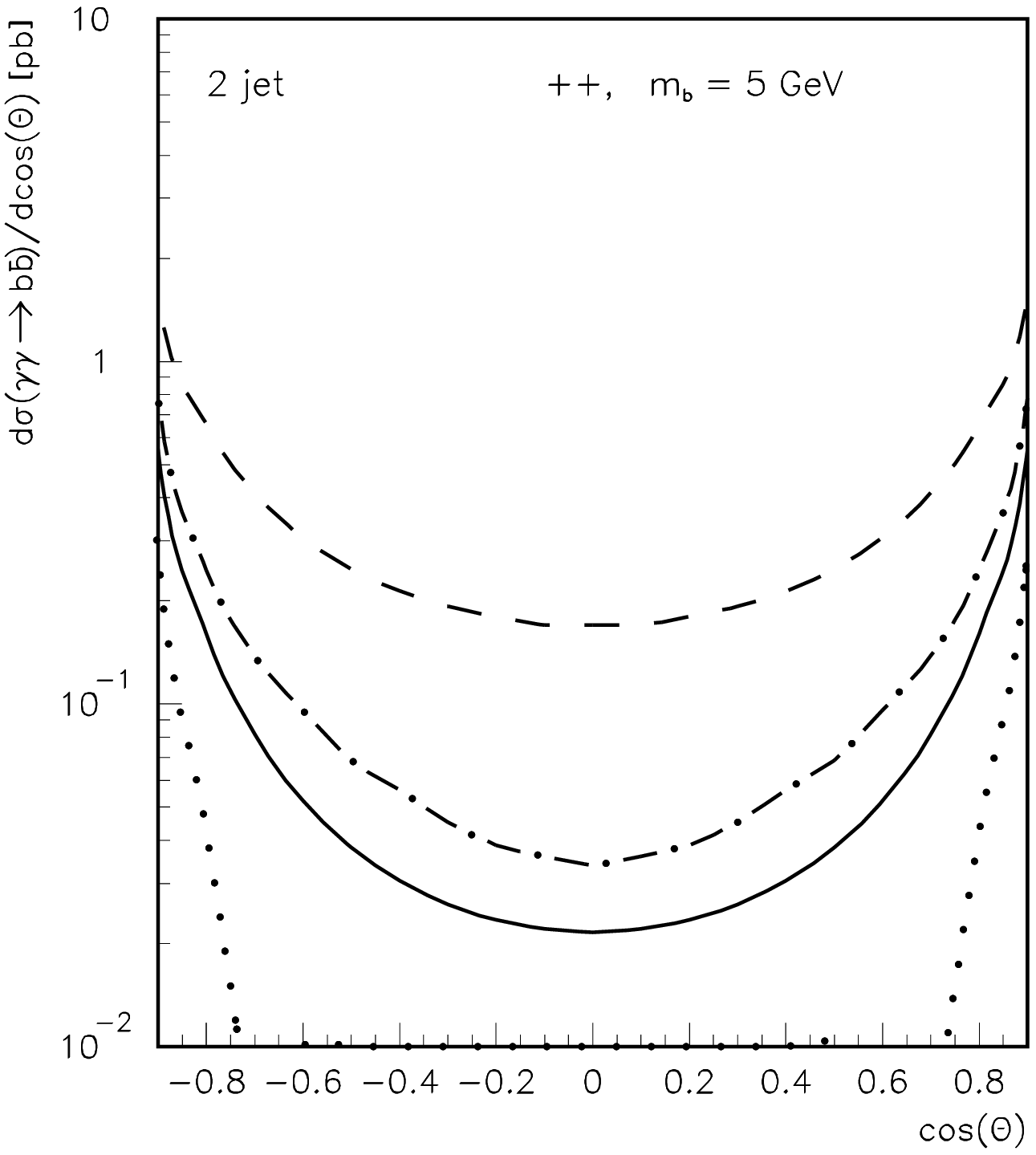,width=7.5cm}}
\put(8,0){\epsfig{file=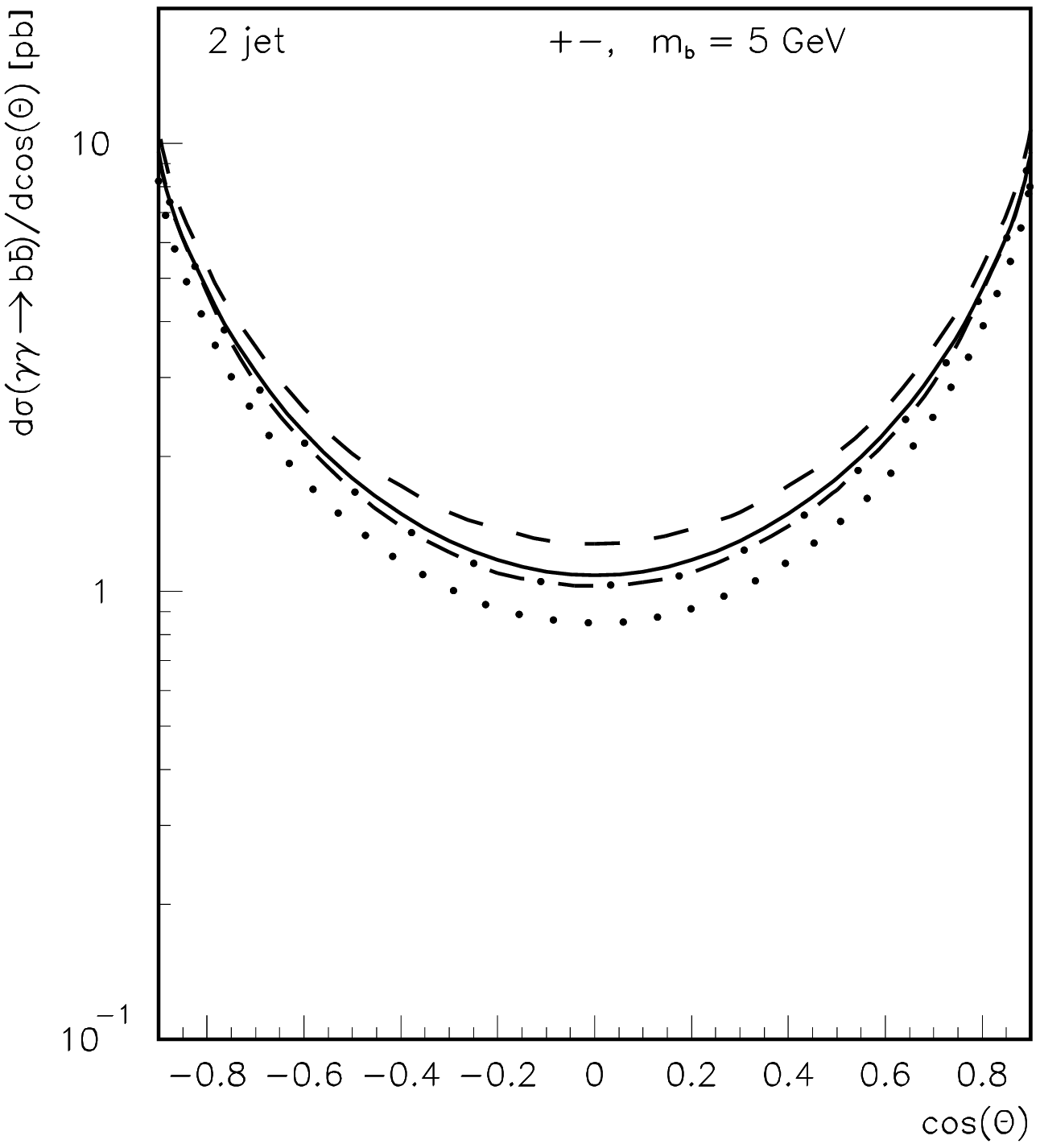,width=7.5cm}}
\end{picture}

\noindent{
Fig.~3. Differential cross sections of the $b\bar b$-pair production versus
b-quark scattering angle at $\sqrt{s}=100$ GeV.  Solid line corresponds to
tree level cross section, dashed line -- total cross section, dash-dotted
line -- two-jet cross section with $y_{cut}=0.08$ and dotted line -- two-jet
cross section with $y_{cut}=0.04$;  (a) cross sections for $J_z=0$ and (b) for
$J_z=\pm2$.
}
\end{figure}

\begin{figure}[t]
\setlength{\unitlength}{1cm}
\begin{picture}(16,7.5)
\put(3,0){\epsfig{file=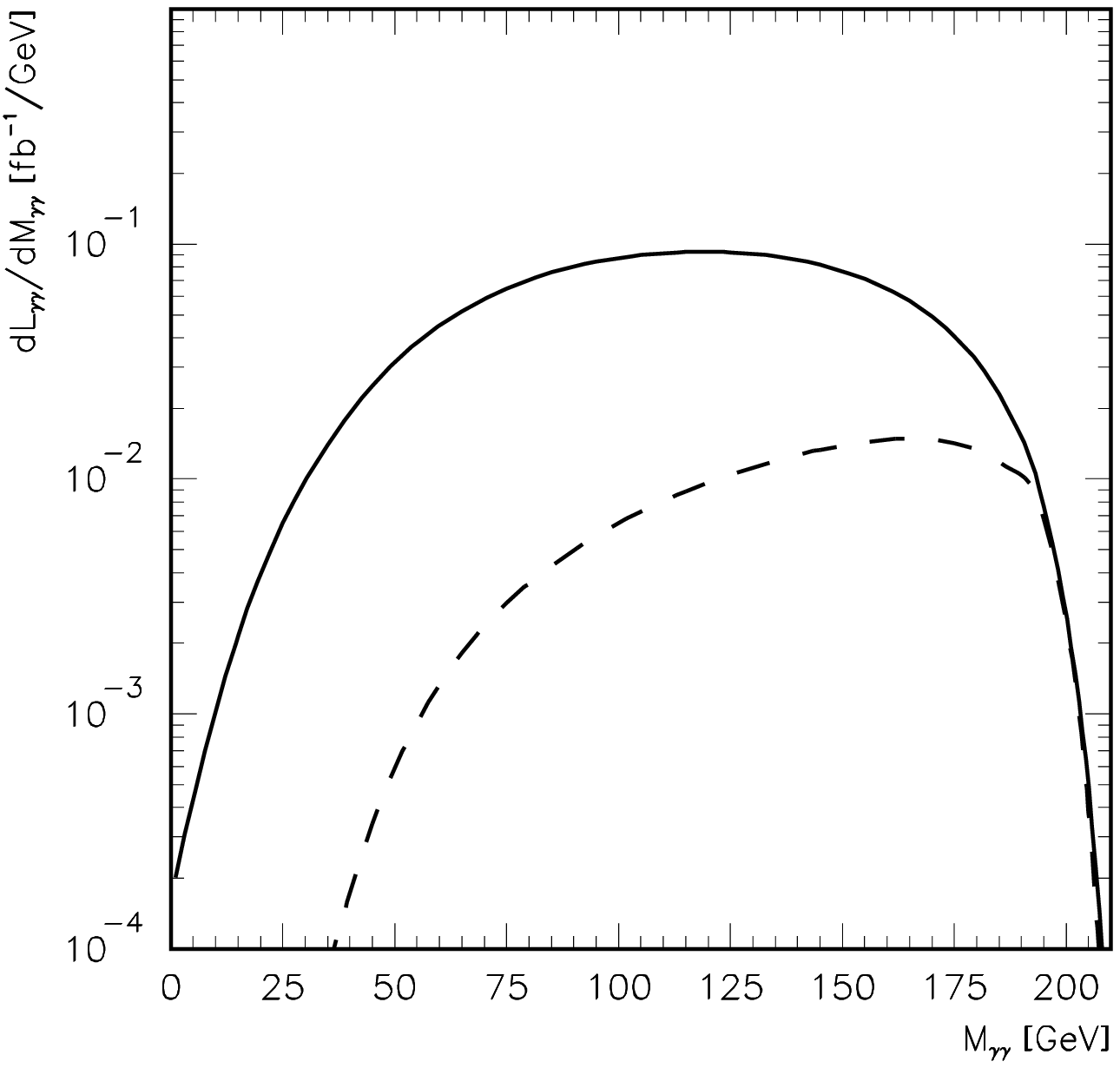,width=7.5cm}}
\end{picture}

\noindent{
Fig.~4. The spread $\g\g$ luminosity distributions to be used in a search for
intermediate-mass Higgs boson. The integrated $e^+e^-$ luminosity of
20~fb$^{-1}$ is assumed.
}
\end{figure}

\begin{figure}[t]
\setlength{\unitlength}{1cm}
\begin{picture}(16,7.5)
\put(0,0){\epsfig{file=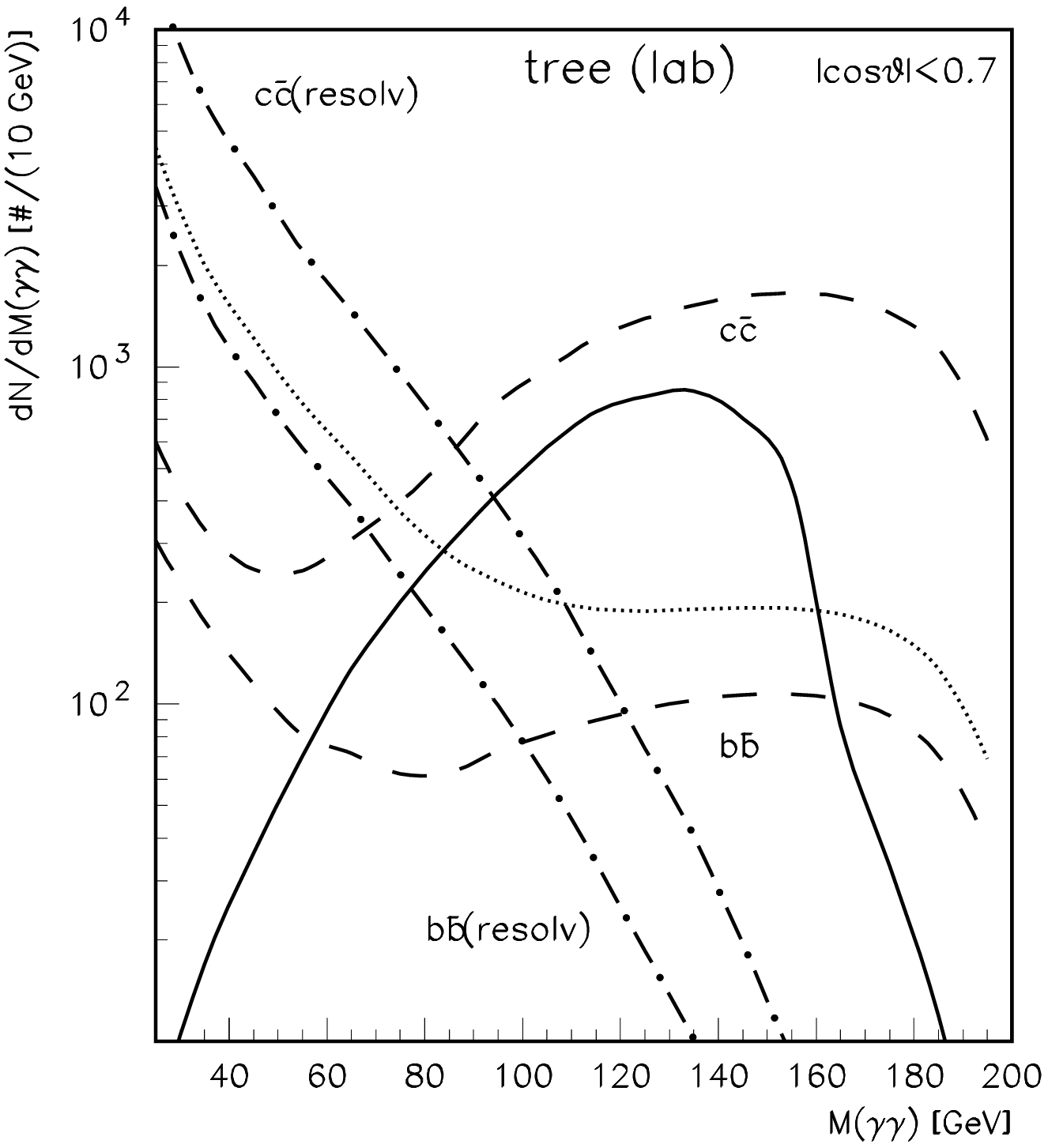,width=7.5cm}}
\put(8,0){\epsfig{file=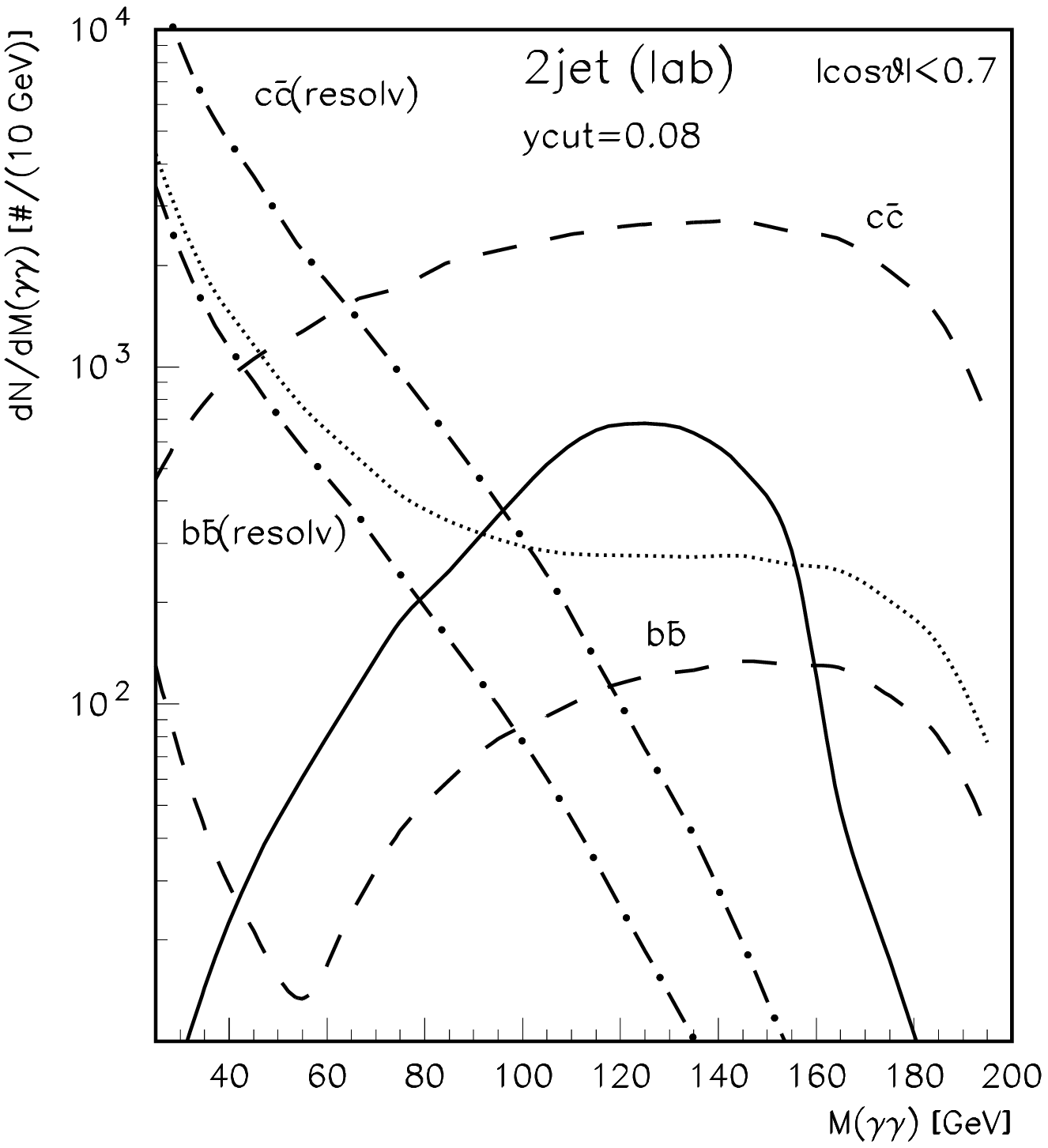,width=7.5cm}}
\end{picture}

\noindent{
Fig.~5. Expected event rates for search for an intermediate-mass Higgs boson in
two jet events. Solid lines correspond to signal events, dashed lines -- to
background events ($\g\g\to~c\bar c$ and $\g\g\to~b\bar b$), dash-dotted lines
-- to resolved photon contributions ($\g\g\to~b\bar b(c\bar c)$); (a) tree
level results and (b) signal  and backgrounds with QCD corrections. Dotted line
in Figure (b) corresponds to the total background ($b\bar b+c\bar c$).
}
\end{figure}

\begin{figure}[t]
\setlength{\unitlength}{1cm}
\begin{picture}(16,7.5)
\put(3,0){\epsfig{file=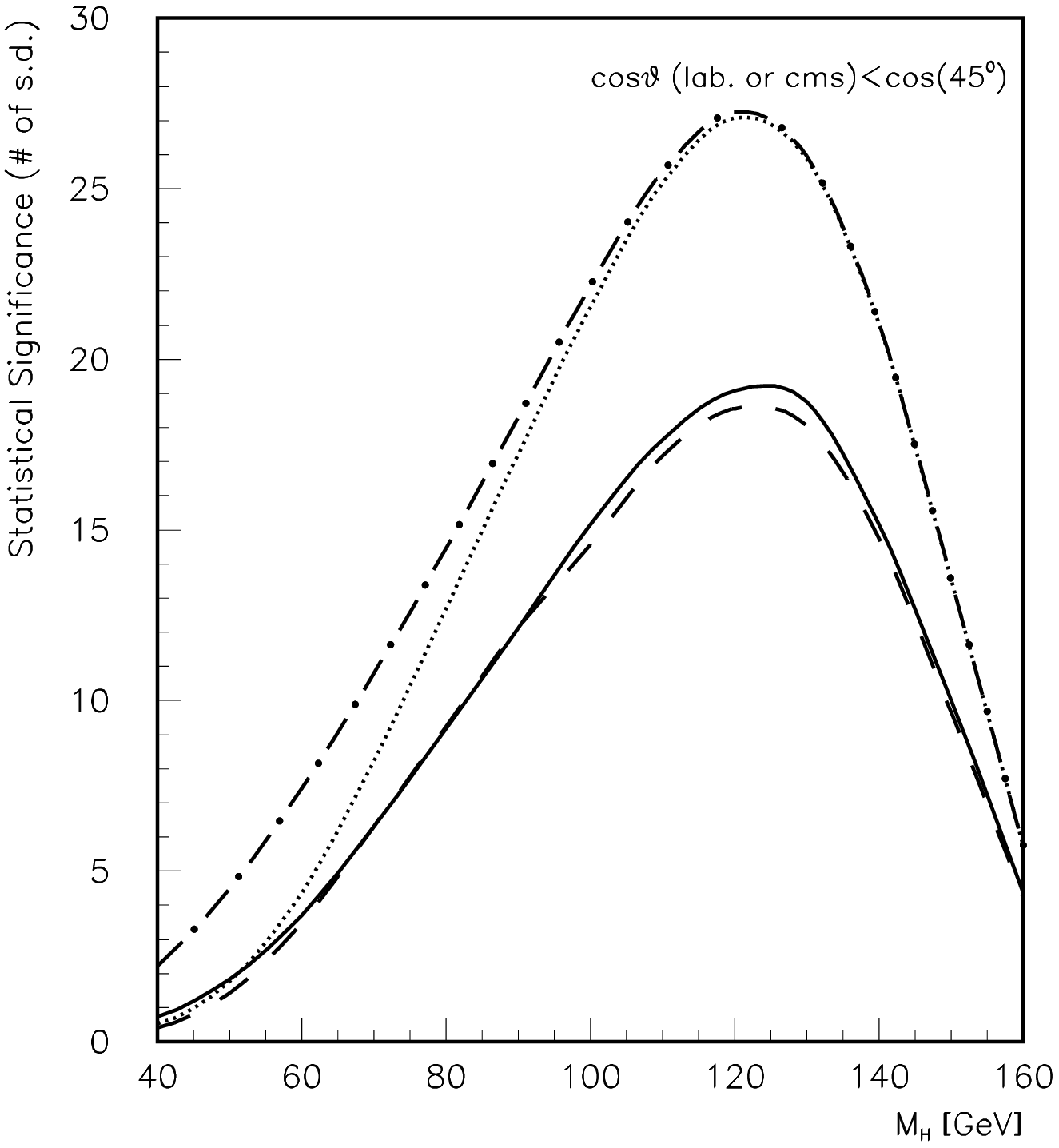,width=7.5cm}}
\end{picture}

\noindent{
Fig.~6. Statistical significance of the Higgs boson signal corresponding to the
event rates in Fig.~5.  Solid line corresponds to the event rates with QCD
corrections (angular cut in the lab. system), dashed line -- to the same ones
with angular cut in the c.m.s., dash-dotted line -- to the tree level rates
without resolved photon contribution \cite{borden}, and dotted line -- to tree
level rates with account of resolved photon contribution.
}
\end{figure}

\begin{figure}[t]
\setlength{\unitlength}{1cm}
\begin{picture}(16,7.5)
\put(3,0){\epsfig{file=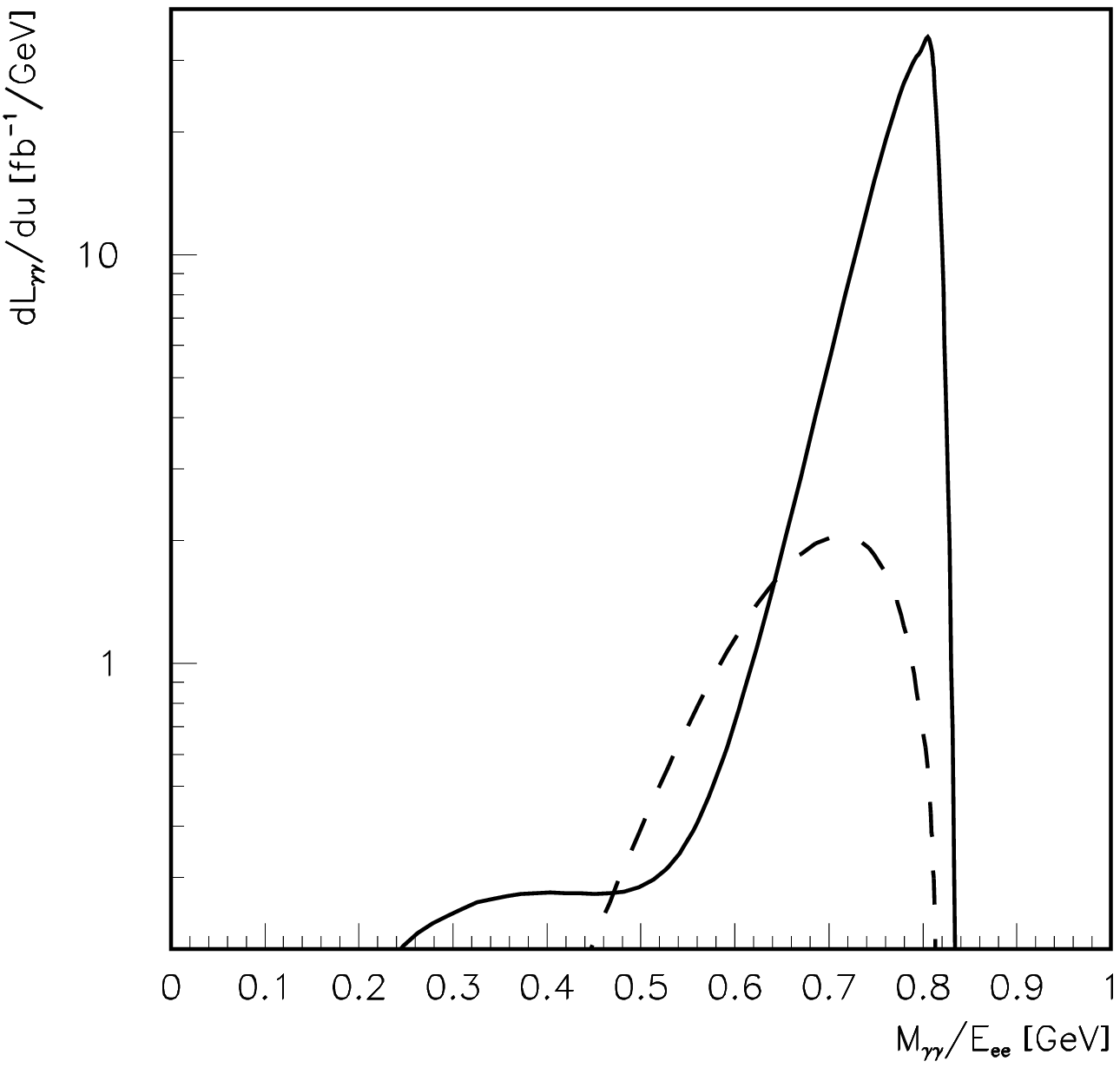,width=7.5cm}}
\end{picture}

\noindent{
Fig.~7. The peaked $\g\g$ luminosity distribution to be used for a precise
measurement of the two-photon width of a Higgs boson.
}
\end{figure}

\newpage
\begin{figure}[t]
\setlength{\unitlength}{1cm}
\begin{picture}(16,7.5)
\put(3,0){\epsfig{file=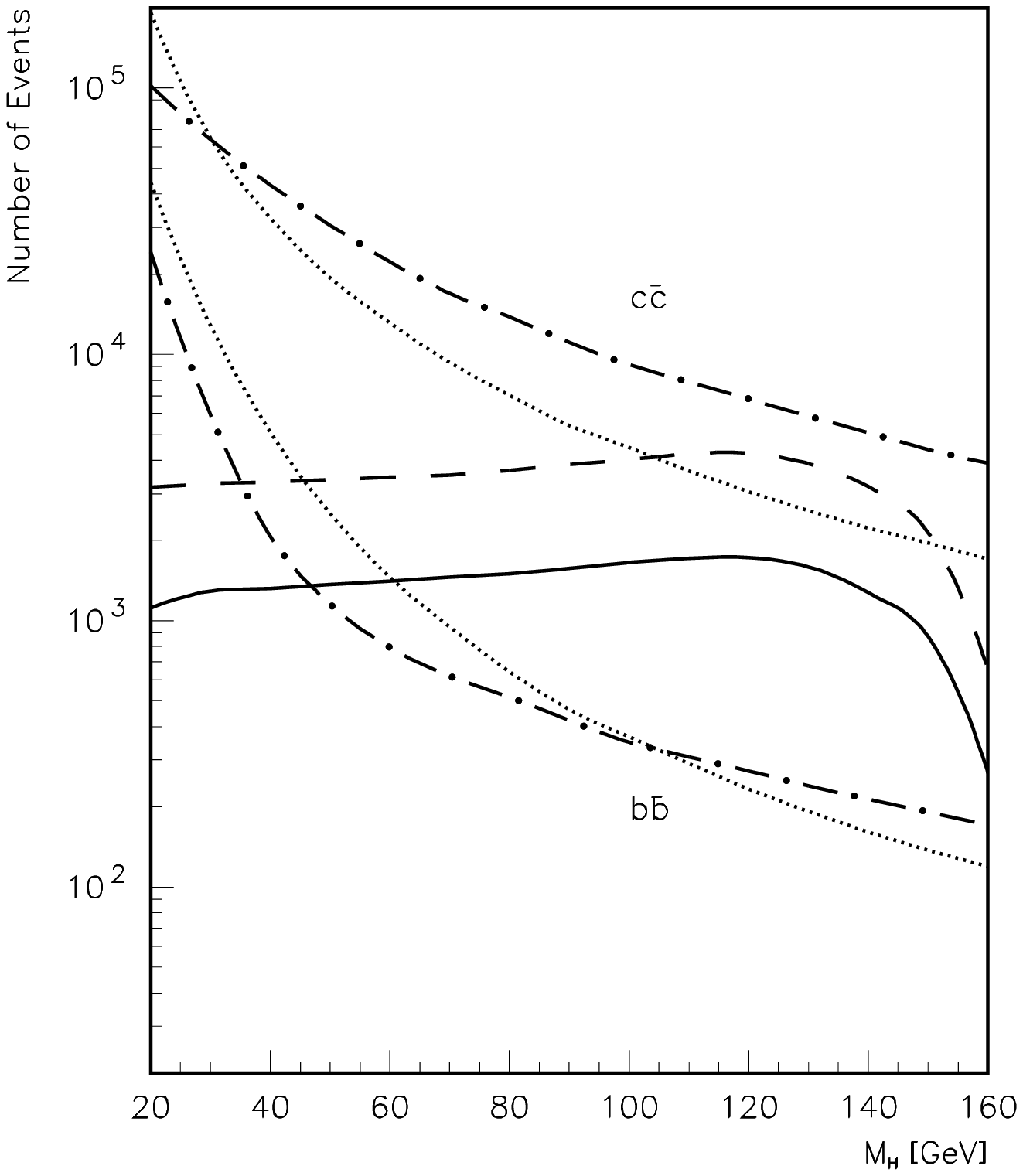,width=7.5cm}}
\end{picture}

\noindent{
Fig.~8. Expected number of the signal and the background events for an
experiment designed to measure the two-photon width of Higgs boson.  Solid
line corresponds to signal events with QCD corrections, dashed line -- to
signal events at tree level, dash-dotted lines -- to background events
with QCD corrections and dotted line -- to background events at tree level.
}
\end{figure}

\begin{figure}[t]
\setlength{\unitlength}{1cm}
\begin{picture}(16,7.5)
\put(3,0){\epsfig{file=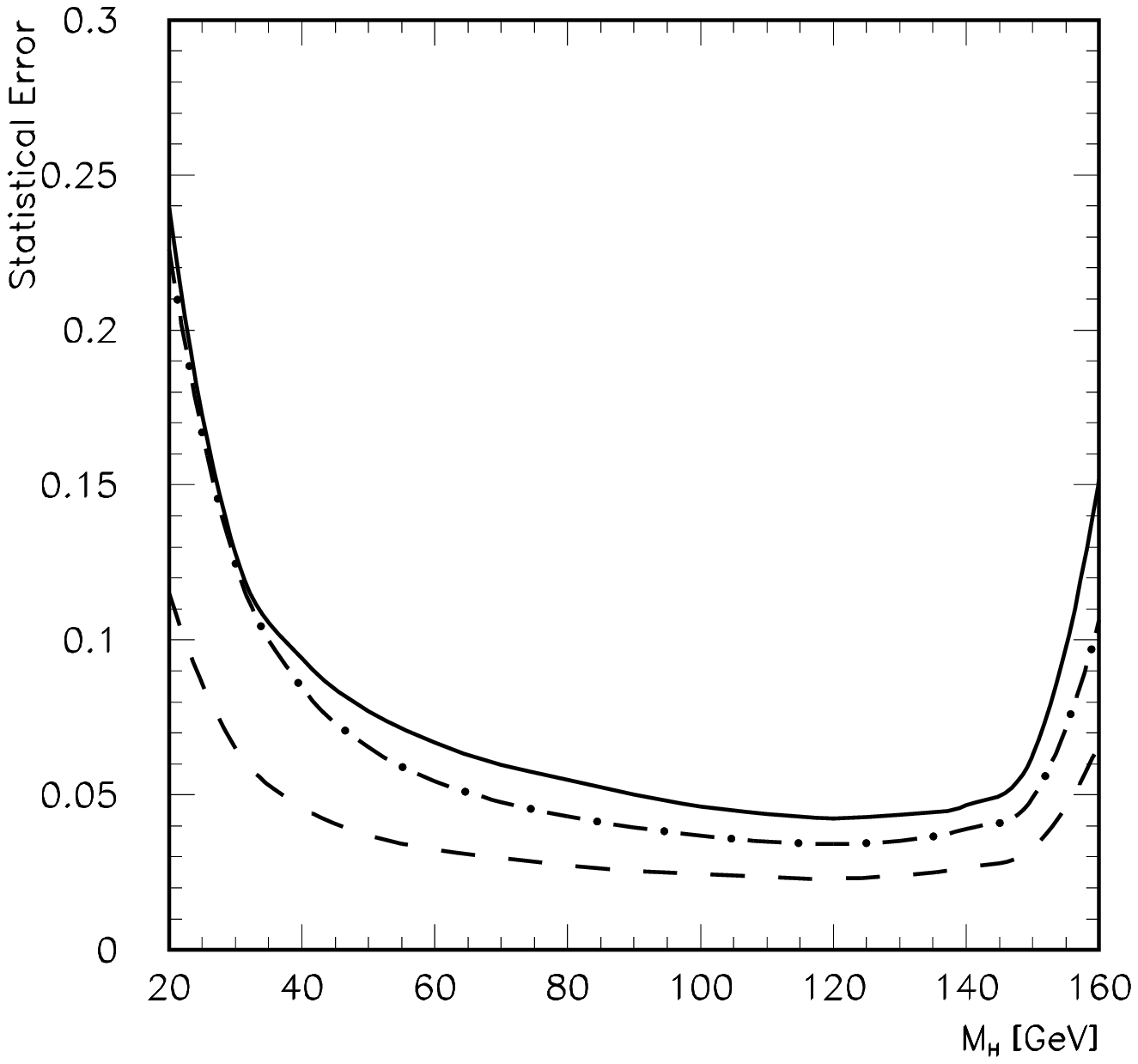,width=7.5cm}}
\end{picture}

\noindent{
Fig.~9. The statistical error in the measurement of the $\Gamma(H\to\g\g)$ for
the event rates given in Fig.~8. Dashed line corresponds to tree level rates
with angular cut in the lab. system, solid line -- to event rates with QCD
corrections and dash-dotted line -- to tree level results from
\cite{borden} (angular cut in the c.m.s.).
}
\end{figure}

\end{document}